\documentclass[11pt]{article}

\usepackage{enumitem} 
\usepackage{xcolor}
\usepackage{authblk}
\usepackage{times}
\usepackage{amsmath, amsthm, amssymb}
\usepackage{color}
\usepackage{graphicx}
\usepackage{cite}
\usepackage{epsfig}
\usepackage{url}
\usepackage{xspace}
\usepackage{references}
\usepackage{hyperfloor}
\usepackage{postponed}

\newif\ifcomments
\commentstrue
 \commentsfalse
\renewcommand{\paragraph}[1]{\vspace{0.2em} \noindent \textbf{#1}}

\proofsfalse

\newcommand{\wasted}{wasted\xspace} 
\newcommand{\nonwasted}{nonwasted\xspace} 
\newcommand{\waste}{waste\xspace} 
\newcommand{\nonwaste}{nonwaste\xspace}

\newcommand{\Waste}{Waste\xspace}

\newcommand{\nc}{\eta} 
\newcommand{\dd}{d}    
\newcommand{\cc}{c}    

\definecolor{light-gray}{gray}{0.95}
\definecolor{gray}{gray}{0.90}

\def\sval{-3pt}
\def\mval{-6pt}
\def\bval{-9pt}

\newcommand{\verysmallskip}{\vspace{.15em}}

\newcommand{\comment}[1]{}

\newcommand{\Cost}{D\xspace}
\newcommand{\Alg}{\textsc{Re-Backoff}\xspace}

\newcommand{\defn}[1]{\textbf{\emph{#1}}}

\newcommand{\myboldmath}{\boldmath}
\renewcommand{\defn}[1]           {{\textit{\textbf{\myboldmath #1\/}}}}

\renewcommand{\epsilon}{\varepsilon}

\definecolor{magenta4}{rgb}{0.5625,0,0.5625}
\definecolor{green4}{rgb}{0,0.5625,0}
\definecolor{orange4}{rgb}{0.98,0.31,0.09}

\newtheorem{theorem}{Theorem}
\newtheorem{corollary}[theorem]{Corollary}

\newtheorem{definition}[theorem]{Definition}

\newtheorem{fact}[theorem]{Fact}

\newcommand{\expect}[1]         {{\rm E}\left[ #1 \right]}

\makeatletter
\renewcommand*{\@fnsymbol}[1]{\ensuremath{\ifcase#1\or \or\dagger\or
    \ddagger \or \mathsection \or \mathparagraph\else\@ctrerr\fi}}
\makeatother
\let\oldfootnoterule\footnoterule
\def\footnoterule{\vskip-1pt\oldfootnoterule\vskip1pt\relax}

\title{How to Scale Exponential Backoff: \\
Constant Throughput, Polylog Access Attempts, and Robustness\\ \vspace{10pt}
\footnote{%
This research was supported in part by
NSF grants
CCF~1114809, 
CCF~1217708, 
CCF~1218188, 
IIS~1247726,  
IIS~1251137,  
CNS~1408695, 
CCF~1439084, 
CNS-1318294, and 
CCF-1420911. 
}
\vspace{-20pt}}

\author[1]{Michael A. Bender}
\author[2]{Jeremy T. Fineman}
\author[3]{Seth Gilbert}
\author[4]{Maxwell Young}

\affil[1]{\small Department of Computer Science, Stony Brook University,
Stony Brook, NY 11794-4400, USA
\hspace{10cm} \texttt{bender@cs.stonybrook.edu}\vspace{5pt}}
\affil[2]{\small Department of Computer Science, Georgetown University, USA\hspace{10cm} \texttt{jfineman@cs.georgetown.edu}\vspace{5pt}}
\affil[3]{\small Department of Computer Science, 
                 National University of Singapore, Singapore\hspace{10cm}
                  \texttt{seth.gilbert@comp.nus.edu.sg}\vspace{5pt}}
\affil[4]{Department of Computer Science and Engineering,  Mississippi State University, USA \hspace{10cm}\texttt{myoung@cse.msstate.edu}\vspace{-15pt}}
\begin{document}

\pagenumbering{gobble}
 
\date{}

\maketitle 


\begin{abstract}
  Randomized exponential backoff is a widely deployed technique for
  coordinating access to a shared resource.  A good backoff protocol
  should, arguably, satisfy three natural properties: (i) it should
  provide constant throughput, wasting as little time as possible;
  (ii) it should require few failed access attempts, minimizing the
  amount of wasted effort; and (iii) it should be robust, continuing
  to work efficiently even if some of the access attempts fail for
  spurious reasons.  Unfortunately, exponential backoff has some
  well-known limitations in two of these areas: it provides poor
  (sub-constant) throughput (in the worst case), and is not robust (to
  resource acquisition failures).

  The goal of this paper is to ``fix'' exponential backoff by making
  it scalable, particularly focusing on the case where processes
  arrive in an on-line, worst-case fashion.  We present a relatively
  simple backoff protocol~\Alg~that has, at its heart, a version of
  exponential backoff.  It guarantees expected constant throughput
  with dynamic process arrivals and requires only an expected
  polylogarithmic number of access attempts per process.

  \Alg is also robust to periods where the shared resource is
  unavailable for a period of time. If it is unavailable for $\Cost$
  time slots, \Alg provides the following guarantees.  When the number
  of packets is a finite $n$, the average expected number of access
  attempts for successfully sending a packet is $O(\log^2( n +
  \Cost))$. In the infinite case, the average expected number of
  access attempts for successfully sending a packet is $O( \log^2(\nc)
  + \log^2(\Cost) )$ where $\nc$ is the maximum number of processes
  that are ever in the system concurrently.

\end{abstract}

\clearpage
 
\pagenumbering{arabic}

\setcounter{page}{1}


\section{Introduction} 
\seclabel{intro}
\vspace{\sval}

\defn{Randomized exponential backoff}~\cite{MetcalfeBo76} is used
throughout computer science to coordinate access to a shared resource.
This mechanism applies when there are multiple processes (or devices,
players, transactions, or packets) attempting to access a single,
shared resource, but only one process can hold the resource at a time.
Randomized backoff is implemented in a broad range of applications
including local-area networks~\cite{MetcalfeBo76} wireless
networks~\cite{kurose:computer, xiao:performance}, transactional
memory~\cite{HerlihyMo93}, lock acquisition~\cite{RajwarGo01}, email retransmission~\cite{Bernstein98,CostalesAl02}, congestion
control (e.g., TCP)~\cite{mondal:removing,jacobson:congestion}, and
a variety of cloud computing
scenarios~\cite{google:gcm,google:best-practices,amazon:error-retries}.

In randomized exponential backoff, when a process needs the resource,
it repeatedly attempts to grab it.  If two processes
\defn{collide}---i.e., they try to grab the resource at the same
time---then the access fails, and each process waits for a randomly
chosen amount of time before retrying. After each collision, a
process's expected waiting time doubles, resulting in reduced
contention and a greater probability of success.

\begin{figure}[h!]

\begin{minipage}[t]{0.96\textwidth}

  \vspace{-5pt}

  \fbox{\colorbox{light-gray}{
      \begin{minipage}{0.96\textwidth}

        \noindent{}In \emph{Exponential Backoff}, when a process needs the resource:

        \begin{itemize}[noitemsep,nolistsep]\renewcommand{\labelitemii}{$\circ$}

        \item Set the time window size $W = 2$.

        \item Repeat  until the resource is acquired:
			
          \begin{itemize}[noitemsep,nolistsep]\renewcommand{\labelitemii}{$\circ$}
			
          \item Randomly choose a slot $t$ in window $W$.  Try to acquire the resource at slot $t$.
			
          \item If the aquisition failed, then: (i) wait until the end of
            $W$, and (ii) set $W = 2W$.
			
          \end{itemize}
			
        \end{itemize}

      \end{minipage}
    } }
            
\end{minipage}

\vspace{-1ex}
\figlabel{backoff}
\end{figure}

Given the prevalence (and elegance) of exponential backoff, it should
not be surprising that myriad papers have studied the theoretical
performance of randomized backoff.  Many of these papers make
queuing-theory assumptions on the arrival of processes needing the
resource~\cite{GoldbergMa96a,GoldbergMaPaSr00,HastadLeRo87,RaghavanUp95,capetanakis:generalized}.
Others assume that all processes arrive in a
batch~\cite{GoldbergJeLeRa93,Gereb-GrausTsa92,GreenbergFlLa87,JACM::GreenbergW1985,Willard84}
or adversarially~\cite{BenderFaHe05,ChlebusKoRo12,ChlebusKoRo06}. What
may be surprising is how many foundational questions about randomized
backoff remain unanswered, or only partially answered:\vspace{3pt}

\begin{itemize}[noitemsep,nolistsep]

\item \emph{Throughput:} It is well known that classical exponential
  backoff achieves sub-constant throughput, in the worst case.  Is it
  possible for an exponential backoff variant to achieve constant
  throughput, particularly in dynamic settings, where arbitrarily
  large bursts of processes may arrive in any time step, leading to
  varying resource contention over time?

\item \emph{Number of attempts:} On average, how many attempts does a
  process make before it successfully acquires the resource?  In
  modifying exponential backoff to achieve constant throughput, is it
  possible to still ensure a small number of unsuccessful attempts?
  These questions make sense in applications where each access attempt
  has a cost.  For example, in a wired network, an unsuccessful
  transmission wastes bandwidth. In a wireless network, an
  unsuccessful transmission wastes energy. In transactional memory, a
  transaction rollback (i.e., an unsuccessful attempt) wastes CPU
  cycles.

\item \emph{Robustness:} How robust is exponential backoff when the
  acquisition process suffers failures? An access attempt could fail
  \emph{even when there is no collision}. Faults could arise due to
  hardware failures, software bugs, or even malicious behavior. For
  example, shared link may suffer from thermal noise, or a
  heavily-loaded server may crash.  A wireless channel may come under
  attack from jamming (see~\cite{xu:feasibility,walters:security}), or
  a server may become unavailable due to a denial-of-service attack
  (DoS)~\cite{hussain:framework}. In transactional memory, failures
  may result from best-effort hardware, since existing hardware
  implementations do not guarantee success even when there are no
  collisions (conflicts) between transactions.

\end{itemize}

\paragraph{The goal: to scale exponential backoff.}
Randomized exponential backoff is ``broken'' in the worst case: it
lacks good throughput guarantees (see, e.g., \cite{BenderFaHe05}) and
is not robust to failures.  While randomized exponential backoff
permits a relatively small number of access attempts, the result is a
subconstant throughput.

The goal of this paper is to ``fix'' randomized exponential
backoff to achieve good {\it asymptotic} performance. 
We modify the protocol as little as possible to maintain its
simplicity, while finding a variant that (1)~delivers constant
throughput, (2)~requires few access attempts, and (3)~works robustly.

\paragraph{Terminology.} Since exponential backoff has many
applications, there are many choices of terminology.  Here we call the
shared resource the ``channel'' and attempts to acquire the resource
``broadcasts.''\footnote{We permit some slight abuses of the English
  language when packets may seem to broadcast themselves.}

\paragraph{Related work.} Exponential backoff is commonly studied in a
network setting, where packets arriving over time are transmitted on a
\defn{multiple-access channel}, and successes occur only if just one
packet broadcasts.

\smallskip

\noindent\emph{Queuing models.}
For many years, most backoff analyses assumed
statistical queueing theory models and focused on finding 
stable packet-arrival rates 
(see~\cite{HastadLeRo87,GoodmanGrMaMa88,RaghavanUp99,
  GoldbergMa96,hastad:analysis,goldberg:contention}).  Interestingly,
even with Poisson arrivals, there are better protocols than binary
exponential backoff, e.g., polynomial backoff~\cite{HastadLeRo87}.
The world runs on exponential backoff; nonetheless, it has long been
known that exponential backoff is broken.

\verysmallskip

\noindent\emph{Worst-case/adversarial arrivals.}
More recently, there has been work on adversarial queueing theory,
looking at the worst-case performance~\cite{GreenbergFlLa87,Willard84,GoldbergJeLeRa97,GoldbergMaRa99,BenderFiGi06,FernandezAntaMoMu13,
  BenderFaHe05,ChlebusKoRo12,ChlebusKoRo06,anantharamu:adversarial-opodis}.
A common theme is that dynamic
arrivals are hard to cope with.  When all the packets begin at the
same time, efficient protocols are
possible~\cite{GoldbergJeLeRa93,Gereb-GrausTsa92,GreenbergFlLa87,JACM::GreenbergW1985,Willard84,FernandezAntaMoMu13}.
When packets begin at different times, the problem is harder.  Dynamic
arrivals has been explicitly studied in the context of the ``wake-up
problem''~\cite{chlebus:better,chlebus:wakeup,chrobak:wakeup}, which
looks at how long it takes for a single transmission to succeed when
packets arrive dynamically.

In contrast, our paper focuses on achieving good bounds for a stream
of packet arrivals (no fixed stations), when all packets must
be transmitted.

\verysmallskip

\noindent\emph{Robustness to wireless interference.}
As the focus on backoff protocols has shifted to include wireless
networks, there has been an increasing emphasis on coping with noise
and interference known as \emph{jamming}. In a surprising
breakthrough, Awerbuch et al.~\cite{awerbuch:jamming} showed that good
throughput is  possible with a small number of access attempts, even if jamming causes disruption
for a constant fraction of the execution.  
A number of elegant results  have
followed~\cite{richa:jamming2,richa:jamming3,richa:jamming4,
ogierman:competitive,richa:efficient-j,richa:competitive-j,KlonowskiDo15}, with good guarantees on throughput and access attempts.

Most of these jamming-resistant protocols do not assume fully dynamic
packet arrivals.  By this, we mean that these protocols are designed
for the setting in which there are a fixed number of ``stations'' that
are continually transmitting packets.  In contrast, we are interested
in the fully dynamic setting, where sometimes there are arbitrarily
large bursts of packets arriving (lots of channel contention) and
other times there are lulls with small handfuls of packets (little
channel contention).

\verysmallskip

\noindent\emph{Relationship to balanced allocations.}
Scalable backoff is closely related to balls-and-bins
games~\cite{BerenbrinkKhSa13,BerenbrinkCzSt06,RichaMi01,
  Vocking03,AzarBrKa99,Mitzenmacher01,BerenbrinkCzEg12,ColeFrMa98}.
Bins correspond to time slots and balls correspond to packets. The
objective is for each ball to land in its own bin; if several balls
share the same bin, they are rethrown. The flow of time gets
modeled by restrictions on when balls get thrown and where they may
land. The results in our paper are ultimately about scalable
randomized algorithms and asymptotic analysis for dealing with bursts
robustly and scalably.

\vspace{\mval}

\vspace{0.5em}
\paragraph{Results.} We devise a ``(R)obust (E)fficient'' backoff protocol, \Alg, that
(1)~delivers constant throughput, (2)~guarantees few failed access
attempts, and (3)~works robustly.  We assume no global broadcast
schedule, shared secrets, or centralized control.

\begin{theorem} \thmlabel{finite}
Let $\Cost$ be the number of slots disrupted by the
  adversary. For a finite number of packets~$n$ injected into the
  system, where $n$ is fixed \emph{a priori}, but not revealed, \Alg 
 guarantees at most an expected constant fraction of \wasted slots 
 (empty slots or slots with collisions)
  and spends $O(\log^2 (n+\Cost))$ access attempts per
  packet, in expectation.
\end{theorem}

\thmref{finite} implies that \Alg delivers constant throughput for
those executions where at least a constant fraction of the slots are
undisrupted:

\begin{corollary}\corlabel{finite}
There exists a constant $c$, such that if $\Cost\leq cn$, 
then \Alg achieves expected constant throughput.  In fact, a stronger property holds: it attains an expected makespan of $O(n)$. \vspace{\sval}
\end{corollary}

An implication of \thmref{finite} is that the number of access
attempts is small relative to the number of disrupted
slots. Specifically: (1) Our protocol is parsimonious with broadcast
attempts in the absence of disruption. (2) If a packet has been in the system for $T$ slots, then it
makes expected $O(\log^2T)$ access attempts, regardless of how many 
 of these $T$ slots were adversarially blocked.

Extending these results to the infinite case, we show that in an
infinite execution, for a countably infinite number of slots, the
protocol achieves both good throughput and few access attempts.\footnote{Achieving
constant throughput in every prefix of an execution is impossible: even in a finite batch setting, it takes $\Omega(\log{n})$
time for the first packet to succeed, yielding at best $\Theta(1/\log{n})$ throughput for that prefix.}
\vspace{\sval}

\begin{theorem} \thmlabel{infinite} 
  For any time $t$, denote by $\Cost_t$ the number of slots disrupted
  before $t$, and by $\nc_t$ the maximum number of packets
  concurrently in the system before~$t$.  Suppose  an infinite
  number of packets get injected into the system.  Then for any
  time $s$, there exists a time $t \geq s$ such that at time $t$,
  \Alg~has at most constant \waste with probability
  1 and expected average $O(\log^2 \nc_t + \log^2 \Cost_t)$ access
  attempts per packet. 
\end{theorem}
Again, this implies that \Alg yields constant throughput in infinite
executions:
\begin{corollary} \corlabel{infinite}
  There exists a constant $c$ such that: for every time $s$ there exists a time $t \geq s$ where if $\Cost_t \leq ct$, 
  then there is constant throughput until time~$t$.
\end{corollary}

\vspace{\bval}

\section{Model: Contention Resolution on a Multiple-Access Channel}

\seclabel{model}
\vspace{\sval}

Time is discretized into \defn{slots} where a process can broadcast a packet, i.e., access the shared resource.  We do not assume a global clock, i.e., there is no universal
numbering scheme for slots.

When there is no transmission (or adversarial disruption, see below)
during a slot, we call that slot \defn{empty}.  A slot
is \defn{full} when one or more packets are broadcast in that
slot. When exactly one packet is broadcast in a slot, that packet
transmits successfully, and the full slot is \defn{successful}.  When
two or more packets are broadcast during the same slot, a
\defn{collision} occurs in this (full) slot.  When there is a
collision, there is ``noise'' on the channel; all packets
transmitting are unsuccessful.
A listening process can determine only whether a slot is full or empty
(but not whether there was a collision).

We assume that a device transmitting a packet can determine whether
its transmission is successful; this is a standard assumption in the
backoff literature (for examples,
see~\cite{hastad:analysis,kwak:performance,goldberg:contention}),
unlike the wireless setting where a full medium access control (MAC)
protocol would address acknowledgments and  other issues. Here,
(as in exponential backoff), we focus
solely on the sending side (backoff component) of the problem.

For simplicity of presentation, we assume there are actually two channels
that processes can use simultaneously: a ``control channel'' and a ``data channel.'' We explain
in \secref{synch-short} and \appref{synch} how to implement our solution using only one channel.

\paragraph{Arbitrary Dynamic  Packet Arrivals.} 
New packets arrive arbitrarily over time.  We do not assume any
bounded arrival rate.  A packet is \defn{live} at any time between its
arrival and its successful transmission. The number of packets in the
system may vary arbitrarily over time, and this number is unknown to
the packets. Without loss of generality, we assume that throughout the
execution of the protocol, there is at least one live packet.  (If
not, simply ignore any slot during which there are no live packets.)

We postulate an \defn{adversary}, who governs two aspects of the
system's dynamics.  (1) The adversary determines the (finite)
number of new packets that arrive in each slot.  (2) The adversary
may arbitrarily \defn{disrupt} slots.  A disruption appears as a
full slot to all packets; any packet transmitted simultaneously
fails.  This model corresponds to collisions on Ethernet or a
$1$-uniform adversary in wireless networks
(see~\cite{richa:jamming2}).

The adversary is adaptive with one exception---the adversary 
decides \emph{a priori} whether the execution  contains infinitely
many packets or a finite number $n$ of packets.  In the finite case,
the adversary chooses the number $n$ \emph{a priori}.  The packets
themselves do not know whether the instance is infinite or finite, and
in the finite case, do not know $n$.  In all other ways, the adversary
is adaptive: it may make all arrival and disruption decisions with
full knowledge of the current and past system state; at the end of a
given slot, the adversary learns everything that has happened in that
slot.

\paragraph{Throughput and \Waste.} We define throughput in the natural way: for an interval $I$,
the throughput $\lambda\in[0,1]$ is the fraction of successful slots in the interval $I$.  (Recall that for the purposes of throughput and waste, we only consider slots when there is at least one packet live in the system.)
  
We also define a notion of ``waste.'' A slot is \defn{\wasted} if
there was a missed opportunity for a successful transmission: the slot
was empty or more than one packet was broadcast.  Otherwise the slot
is \defn{\nonwasted}, i.e., successful or disrupted.  A disrupted slot
is not seen as \wasted, since it could never be used for a successful
packet transmission.  The \defn{\nonwaste} $\Lambda \in[0,1]$ of an
interval $I$ is the fraction of \nonwasted slots in $I$, and the
\defn{\waste} is $1-\Lambda$.  In the absence of disruption,
throughput and \nonwaste are identical.

\begin{definition} \deflabel{throughput} Consider interval
  $\mathcal{I}$ having $N_{\mathcal{I}}$ successful transmissions and
  $\Cost_{\mathcal{I}}$ disrupted slots.  The \defn{throughput} of
  $\mathcal{I}$ is $\lambda= N_{\mathcal{I}} /|\mathcal{I}|$, the
  \defn{\nonwaste} is $\Lambda=(N_{\mathcal{I}} +
  \Cost_{\mathcal{I}})/|\mathcal{I}|$, and the \defn{\waste} is
  $1-\Lambda$.  
  The throughput/waste of a finite execution is the throughput/waste for the interval $[0,T]$, where $T$ is the latest any packet completes.
\deflabel{throughput-interval}
\end{definition}

\vspace{-2em}

\begin{definition}
  An infinite instance has $\Lambda$-\nonwaste if, for any slot $t$,
  there exists a slot $t'\geq t$ where interval $[0,t']$ has $\Lambda$
  \nonwaste. An infinite instance has $\lambda$-throughput if, for any
  slot $t$, there exists a slot $t'\geq t$ where interval $[0,t']$ has
  $\lambda$ throughput.  \deflabel{throughput-infinite}
\end{definition}
\vspace{\sval}

The throughput/\nonwaste does not depend on the arrival rate, even with no
restrictions on arrivals.  The arrival rate could be higher than
feasible for an arbitrary period of time (e.g., two packets arrive
every slot), and the system continues to deliver good throughput
(even as the number of backlogged packets necessarily grows).

There are also no restrictions on the distribution of disruptions.
The adversary can choose to disrupt arbitrarily large intervals of
slots.  When there are enough nondisrupted slots, constant throughput
resumes.

\vspace{\mval}

\section{Algorithm}
\seclabel{algorithm}
\vspace{\sval}

\begin{figure}[t]
    \begin{center}
	
        \fbox{\colorbox{light-gray}{     
		
        \begin{minipage}[t]{0.96\textwidth}

        \fbox{\colorbox{gray}{

	\begin{minipage}{0.96\textwidth}

	\noindent{}\Alg~for a node $u$ that has been active for $s_u$ slots \vspace{-7pt}\

\begin{itemize}\renewcommand{\labelitemii}{$\circ$}

                          \vspace{-.3em}
			\item With probability $\frac{c\max\{\ln s_u,~1\}}{s_u}$, send busy tone on the control channel. 
                          \vspace{-.7em}
			\item With probability $\frac{d}{s_u}$ send $m_u$ on the data channel and, if successful, then terminate. 
                          \vspace{-.7em}
			\item Monitor the data channel.  If at least a
                          $\gamma=15/16$-fraction of data slots have
                          been empty since node $u$ became active,
                          then become inactive.  
	\end{itemize}
	\end{minipage}
	}
	}	
	           
	\fbox{\colorbox{gray}{
	
	\begin{minipage}{0.96\textwidth}
	\noindent{}\Alg~for an inactive node \vspace{-7pt}\

	\begin{itemize}\renewcommand{\labelitemii}{$\circ$}
                          \vspace{-.5em}
	\item Monitor each control slot. If a slot is empty, then become active next slot. \vspace{-0pt}
		\end{itemize}	
\end{minipage}
}
}

	\end{minipage}

	}
	}
    \end{center}

\vspace{-18pt}
\caption{Pseudocode for \Alg.}\figlabel{pseudocode}\vspace{-8pt}
\end{figure}

This section presents our backoff protocol. To simplify the presentation, we assume throughout that there are two
communication channels, a \defn{data channel}, on which packets are
broadcast, and a \defn{control channel}, on which a ``busy signal'' is
broadcast.  See \secref{synch-short} how to implement this algorithm
on one channel.

For a given packet
$u$, let $s_u$ be the number of slots it has been active for.  Our
protocol has the following structure (see \figref{pseudocode} for
pseudocode):\vspace{\mval}
 \begin{itemize}[noitemsep]
 \item Initially, each packet is \defn{inactive}; it makes no attempt
   to broadcast on either channel.

 \item Inactive packets monitor the control channel.  As soon as the
   packet observes an empty slot on the control channel, it becomes
   \defn{active}.

 \item In every time slot, an active packet broadcasts on the data
   channel with probability proportional to how long it has been
   active, i.e., packet $u$ broadcasts with probability $d/s_u$, for a
   constant $d = 1/2$.  It also broadcasts on the control channel
   with probability $c\max(\ln s_u, 1)/s_u$, for a constant $c >0$.

 \item A packet remains active until  it transmits
   successfully or  sees too many empty slots.  Specifically,
   if packet $u$ has observed $\gamma s_u$ empty slots,
    the packet reverts to an \defn{inactive} state, and the
   process repeats.
\end{itemize}

\vspace{\mval}In essence, our protocol wraps exponential backoff with
a coordination mechanism (i.e., the busy channel) to limit entry, and
with an abort mechanism to prevent overshooting.  In between, it runs
something akin to classical exponential backoff (instantiated by
broadcsting in round $t$ with probability $1/t$, instead of using
windows).  One aspect that we find interesting is how little it takes
to fix exponential backoff.

\vspace{\mval}

\section{Protocol Design}
\vspace{\sval}

This section gives the intuition behind the design of \Alg.

Consider the following simple protocol that \Alg builds upon. Packets are
initially inactive.  Whenever there are no active packets, all packets
in the system become active and run an asymptotically optimal
\defn{batch} backoff protocol on the data channel (e.g., SawTooth
Backoff~\cite{BenderFaHe05}).  Active packets \emph{all} broadcast a \defn{busy
  tone} on every control-channel slot, and inactive packets wait for
silence on the control channel.\footnote{Busy tones are also used in
  mutual exclusion and MAC protocols (see,
  e.g.,~\cite{haas:dual,wu:receiver}) for coping with hidden terminal
  effects.}  The busy tone contains {\it no data}, and it serves only to prevent newcomers from activating
until all currently active packets have transmitted successfully. 

The busy tone yields a \defn{batch invariant}: there is only one batch
running in the system at a time, which allows the throughput
guarantees of the batch protocol to extend to arbitrary arrivals.
Unfortunately, this basic protocol yields an unacceptable number of
access attempts---one attempt per active packet per time step due to
the busy tone.  But even this primordial protocol is interesting because it shows a simple strategy for achieving constant throughput,
in contrast to classical exponential backoff; see
\figref{fig-exp-backoff-struggles}.

We require a cheaper busy tone. A natural approach is for active
packets to broadcast randomly on the control channel. 
This modified protocol broadcasts less, but it
suffers the occasional \defn{control failure}, where the busy tone
disappears even though some packets are still active.

The question is how active packets should respond to control
failures. A plausible approach is to reset \emph{every} packet, making
every packet in the system restart in a single new batch. With no
disruptions, this new protocol achieves constant throughput with a
polylogarithmic number of access attempts.  

But it is not robust to disruptions.  The adversary has too much
control: it can spoof the busy tone until packets have backed off a
lot.  The adversary then stops, and now packets have a very low
probability of making an access attempt before a control failure
causes a reset, which forces packets to join a new batch.  Using this
strategy, the adversary can prevent almost all of the packets in each
batch from broadcasting successfully, forcing them to reset many
times.  Specifically, the adversary can keep packets in the system for
$T\gg n$ time steps, and it can force $T^{\Theta(1)}$ access attempts
rather than $\mbox{polylog}(n+T)$, access attempts, as specified by
\thmreftwo{finite}{infinite}.

\Alg~addresses the previous concern by avoiding immediate resets; a
packet resets only once a constant fraction of slots during its
current batch are empty. Intuitively, the reset condition means that
any packet that was reset could easily have chosen one of the empty
slots, and just got unlucky.  Any packet that enters a batch has at
least a constant probability of broadcasting successfully in that
batch and at most a constant probability of resetting.  Therefore, in
\Alg~a packet joins an expected constant number of batches before
succeeding.

And so \Alg~sacrifices the batch invariant; multiple batches may exist in the
system simultaneously and, consequently, we have put the throughput
guarantee in jeopardy.  This is because batch protocols do not perform
well with dynamic arrivals. Even exponential backoff, which is already
suboptimal on batches, performs asymptotically worse under dynamic
arrivals.

The probabilistic busy tone and delayed reset serve together as a
``leaky-mutual-exclusion'' protocol, which keeps out many overlapping
batches, but allows others to ``leak'' into the critical
section. (This contrasts with the (error-free) busy tone and
aggressive reset mechanisms, each of which deterministically ensures
mutual exclusion.)  Most of the technical contribution of our paper
has to do with proving that \Alg still guarantees constant contention a
constant fraction of the time, and therefore ensures constant
\nonwaste (and therefore constant throughput
when a constant fraction of slots are jammed), 
despite leaky mutual exclusion.  The idea is to prove that
there are enough prefixes of slots so that: if the contention (the sum
of broadcast probabilities) is much more or much less than a constant
for $X$ slots, then there are $\Omega(X)$ slots where contention is
$\Theta(1)$ and so many packets should succeed.

Digging deeper, the technical hurdle that contention arguments seem to
have is that contention changes over time in ways that are hard to
characterize. For example, if the contention in a given time slot
comes from a small number of young packets, then it will drop quickly
over time (unless another batch activates), whereas if the contention
comes from a large number of older packets, then it will drop 
gradually. Thus, there is a funny and unpredictable way in which the
contention changes as a function of time.

Besides its complexity, what makes this proof unusual to us is that we
are deprived of some of our favorite tools: high-probability arguments
e.g., using Chernoff bounds.  This tool is denied to us because the
bursts may be arbitrarily smaller than the number of packets ever to
enter the system.

Ultimately, we have a rather simple protocol that maintains, at its
core, exponential backoff---while at the same time delivering the
three desirable properties: constant throughput, few attempts, and
robustness.

\vspace{\mval}

\section{Throughput and \Waste Analysis}\vspace{\sval}
\seclabel{analysis}

In this section, we analyze the throughput of the~\Alg~protocol, showing that it achieves at most constant \waste in both the finite and infinite cases.  
All omitted proofs appear in \appref{proofs}. 
 
Let $s^t_j$ be the age of packet $j$ in slot $t$, i.e.,
the number of slots that it has been active.    At time $t$, we define the \defn{contention} to be
$X(t) = \sum_i 1/s^t_j$, where we sum over all the active packets. 
Thus, the expected number of broadcasts on the data channel in slot $t$ is $\dd X(t)$. We divide the packets that are active in slot $t$ into \defn{young} and
\defn{old} packets.  For every slot $t$, we define the value $\sigma_t$ to be the minimum age out of all active packets such that the following hold:~~(i)~$\sum_{i : 0 < s^t_i \leq \sigma_t} 1/s^t_i \geq
X(t)/2$;  (ii)~$\sum_{i : s^t_i \geq \sigma_t} 1/s^t_i \geq X(t)/2$.  That is, active packets with age $\leq \sigma_t$ have at least half
the contention, and active packets with age $\geq \sigma_t$ have at
least half the contention.  We call these two sets the \defn{young} and \defn{old} nodes, respectively. Note that packets with age exactly $\sigma_t$ are \emph{both} young and old.  \vspace{\sval}

\LemmaProofLater{sk}{
For all times $t$, $\sigma_t$ is well-defined.\vspace{\sval}
}
{
Sort the packets by age so that $s_1 \leq s_2 \leq \ldots$.  Let $k$ be the minimum index such that $X(t)/2 \leq \sum_{i=1}^k 1/s_i\leq X(t)/2 + 1/s_k$.  
Let $\sigma_t = s_k$.  Notice that the young packets have contention at least $X(t)/2$, and the old packets have contention at least $X(t)/2$.
}
\vspace{\sval}

\noindent 
We say that a \defn{control failure} occurs in slot $t$ if no node broadcasts on the control channel during the slot.  Recall that (1) a packet activation can occur only immediately after a control failure and (2) a packet $j$ \defn{resets}  at time $t$ if $t$ is the first slot during 
$j$'s  lifetime  $[t-s^t_j, t]$ of $s_j$ slots, for which at least $\gamma s_j$ slots are empty.

\paragraph{Overview.} In \secref{individual}, we relate performance to contention.  The tricky part is to bound \emph{how often} and \emph{for how long} the contention stays high.  In \secref{epochs}, we break the execution up into epochs, structuring the changes in contention.  We can then analyze the control failures (\secref{control}) and resets (\secref{resets}) as a function of contention.  This leads to a key result (\corref{epochsgood}) in \secref{streaks} that shows that an epoch is ``good'' (in some sense) with constant probability.  

One tricky aspect remains: the adversary may use the results from earlier epochs to bias later epochs by injecting new packets at just the wrong time.  We introduce a simple probabilistic game, \emph{the bad borrower game}, to capture this behavior and show that it cannot cause  much harm (in Sections~\ref{sec:badborrower}--\ref{sec:infinite}).  Finally, we assemble the pieces in Sections~\ref{subsec:interstitial-finite} and~\ref{subsec:interstitial-infinite}, showing that we achieve at most a constant-factor 
\waste. 

\vspace{\mval}
\subsection{Individual Slot Calculations}\vspace{\sval}
\seclabel{individual}

The next two lemmas look at the probability of a broadcast as a function of the contention (assuming $\dd \leq 1/2$), first looking at successful broadcasts and then all broadcasts---even those that result in a collision.
 
\LemmaProofLater{efficiency}{
  For a given slot $t$ in which there is no disruption, the
  probability that some packet successfully broadcasts at time $t$ is
  at least $\frac{\dd\,X(t)}{e^{2\dd X(t)}}$. \vspace{\sval} 
}
{
  Packet $j$ is successful with probability
  $\frac{\dd}{s_j}\prod_{i\neq j} (1-\frac{\dd}{s_i})$.  At most one
  packet is successful, so the success events for each node are
  disjoint. The probability that some packet succeeds is thus at least
  $\frac{\dd}{s_1} \prod_{i}(1-\frac{\dd}{s_i}) + \frac{\dd}{s_2}
  \prod_{i}(1-\frac{\dd}{s_i}) + \cdots + \frac{\dd}{s_k}
  \prod_{i}(1-\frac{\dd}{s_i}) =
  \left(\prod_{i}(1-\frac{\dd}{s_i})\right)\cdot \sum_j\frac{\dd}{s_j}
  \geq \frac{\dd X(t)}{e^{2 \dd X(t)}}$.  The denominator follows from
  the fact that $1-x \geq e^{-2x}$ for $0\leq x \leq 1/2$, and hence
  $\prod_i (1-\frac{\dd}{s_i}) \geq e^{-2d\sum_i(1/s_i)}$.
}

\LemmaProofLater{nonempty}{
  The probability that some packet is broadcast (not necessarily
  successfully) in slot $t$ is at least $1-e^{-d\,X(t)}$ and at most
  $1- e^{-2d\,X(t)}$.  The probability of a collision in the slot is
  at most $(1- e^{-2d\,X(t)})^2$.\vspace{\mval}  
}
{
  The probability that no nodes broadcast is $\prod_i\left(1 -
    \frac{d}{s_i}\right) \leq e^{-d\,X(t)}$. Conversely,
  $\prod_i\left(1 - \frac{d}{s_i}\right) \geq e^{-2d\,X(t)}$ by the
  fact that $1-x \geq e^{-2x}$ for $0\leq x \leq 1/2$.
The probability of a collisions is at most the square of the probability of a given broadcast, because this is the probability you get if we allow each packet to broadcast twice. 
}

\subsection{Epochs, Streaks, and Interstitial Slots}\vspace{\sval}
\seclabel{epochs}

An execution is divided into two types of periods: 
\defn{epochs} and \defn{interstitial slots}.\vspace{\sval}  

\begin{definition}
  Each time $t_0$ when a packet is activated, a new \defn{epoch}
  begins (and any earlier epoch ends).   When an epoch ends, either a new epoch
  begins (if a new packet is activated) or there is a gap between epochs 
  called the \defn{interstitial slots}.   To
  describe the duration of an epoch, we have two cases. 

  If the contention is not too high at the start of an epoch,
  specifically, if $X(t_0) < 8$, then the epoch consists of a
  single timestep $t_0$.  We call such an epoch a \defn{unit epoch}.
  
  If $X(t_0) \geq 8$, then the epoch is subdivided further into a
  sequence of \defn{streaks}, with the first streak beginning at $t_0$.  If a streak begins at time $t$, then it ends at time
  $t' = t+\sigma_t$ (or at the start of a new epoch, whichever occurs
  first).  If $X(t') < 8$, then the epoch ends.  Otherwise another
  streak begins at time $t'$ and ends at time $t'+\sigma_{t'}$.  \vspace{\sval}
\end{definition}

\noindent In general, we say that an epoch is \defn{disrupted} if at least 1/4 of its slots are disrupted.  We next bound the change in contention during a streak. \vspace{\sval}

\LemmaProofLater{contentionupper}{
Assume that some streak begins at time $t$ and that no control failures occur during the streak.  Then $X(t+\sigma_t) \leq 3X(t)/4$. 
}
{
During the streak, all the young packets at time $t$ \emph{at least} double in age (since they each have age at most $\sigma_t$), leading their contention to at least halve.  Moreover, the young packets at time $t$ have contention at least $X(t)/2$, so the total contention reduces by at least $X(t)/4$.  Since there are no control failures, there are no new packets activated and hence no increase in contention.
}

\LemmaProofLater{contentionlower}{
Assume that some streak begins at time $t$, where $X(t) \geq 8$, and that no resets occur during the streak.  Then for all $t' \in [t, t+\sigma_t]$, $X(t') \geq X(t) / 8$. \vspace{\mval}
}
{
Since there are no resets (and no activations, by definition) during the streak, the contention only decreases due to packets completing and due to  increasing age.  Consider the old packets at time $t$ (which have contention at least $X(t)/ 2$ at time $t$).  Since each of these packets at most doubles in age (since they have age $\geq \sigma_t$), their total contention remains at least $X(t)/4$ throughout the streak.

Some of these packets may finish, thus reducing the contention further.  Assume that the old packet with the largest contention completes in every slot---notice that each such packet that finishes reduces the contention by at most $1/\sigma_t$.  Thus, if one such packet finishes in each slot of the streak, the total contention is reduced by at most $\sigma_t / \sigma_t = 1$.  Thus, throughout the streak, the total contention remains at least $X(t)/4 - 1 \geq X(t)/8$ (since $X(t) \geq 8$). 
}

\subsection{Control Failures}\vspace{\sval} 
\seclabel{control}

\noindent Next we look at the probability of a control failure as a
function of contention.  The next lemma argues that for the next slot
in a streak (that has not yet had any failures), the old packets
provide enough contention to make a control failure in the
next slot is unlikely. The subsequent lemma takes a union bound over
all slots in the streak to conclude that it is unlikely for any
control failure to occur in the streak. 

\LemmaProofLater{singleslot}{
For a fixed time $t$ when a streak begins, consider a control slot at time $t'$ during the streak.  Assume that there are no control failures or resets during the streak prior to time $t'$.  Then the probability of a control failure in slot $t'$ is at most $\left(\sigma_t\right)^{-\frac{\cc X(t)}{2}}$.
}
{
The probability of a control failure in slot $t'$ is at most: 
\begin{eqnarray*}
\prod_i \left(1-\frac{\cc\ln s^{t'}_i}{s^{t'}_i}\right) & \leq & e^{-\cc \sum_i \frac{1}{s^{t'}_i}\ln s^{t'}_i}\\
& = & \sigma_t^{-\cc \sum_i \frac{\ln s^{t'}_{i}}{s^{t'}_i\ln \sigma_{i}}}\\
& \leq & \sigma_t^{-\cc \sum_{\text{``old'' $i$}} \frac{1}{s^{t'}_i}}\\
& \leq & \sigma_t^{-\cc X(t)/2} 
\end{eqnarray*}
\noindent where the probability on the last line is a function of $t$ (not $t'$) by \lemref{sk}.
}

\LemmaProofLater{noinject}{
For a fixed time $t$, consider a streak beginning at time $t$.  Assume that no reset occurs during the streak.  The probability that a control failure occurs in the interval $[t, t+\sigma_t]$ is at most $(\sigma_t)^{-b  X(t)}$ for a constant $b>0$ depending only on constant $\cc$ in our algorithm.\vspace{\sval}
}
{
Assuming there are no control failures during time $[t, t+s-1]$, the probability of a control failure at time $t+s$  is at most $\left(\sigma_t\right)^{-\frac{\cc X(t)}{2}}$.  Taking a union bound over the $\sigma_t$ time slots, the probability of a control failure happening in any time slot is at most $\left({\sigma_t}\right)^{-\frac{\cc X(t)}{2}+1}$.
}

\subsection{Bounding Resets}\vspace{\sval}
\seclabel{resets}

We next bound the probability that a reset takes place during an epoch.  We show that with constant probability, a packet does not reset during an epoch; this is true since for any prefix of the epoch, there are sufficiently many broadcasts to prevent a reset.  We first look at an abstract coin flipping game:

\LemmaProofLater{trialsuccesses}{
  Consider a sequence of Bernoulli trials each with probability $p$ of success. If the first $16/p$  
  trials are all successful with probability at least $q$, then with probability at least $q/2$, for
  all $i$, the first $i$ trials contain at least $ip/4$ successes.
}
{  
  Break up the trials into geometrically increasing subsequences of
  $2^k$ trials each.  We say that a ``failure'' occurs in the $k$th
  subsequence if there are fewer than $p2^k/2$ successful trials within
  that subsequence.  Using a Chernoff bound, the failure probability is
  at most $e^{-p2^k/8}$.  Using a union bound, the probability of any
  failure for subsequence $k \geq \lg(1/p)+4$ is at most $\sum_{k=\lg(1/p)+4}^\infty e^{-p2^k/8} =
  \sum_{j=1}^\infty e^{-2^j} \leq 1/2$. Therefore, with probability at least $q\cdot (1/2)$, the first $16/p$  
  trials are a success \emph{and} every subsequence has at least $p2^k/2$ successes.
  
  Now, suppose there is no failure in any subsequence, i.e., each has at
  least $p2^k/2$ successful trials.  Pick any cutoff point in the subsequence of size $2^i$ and examine the total of $2^{i+1}$ trials up to this point. The previous subsequences for $k=1, \ldots, i-1$ each contain at least $p2^{k}/2$ successful trials, for a total of at least $p\cdot 2^{i}/2$ successful trials. Therefore, up to the cutoff point,  at least a $p/4$-fraction of the trials are successful.
}

\noindent We now conclude in the next lemma that a reset occurs during an epoch with a bounded constant probability.  

\LemmaProofLater{epochrestart}{
Consider an epoch that begins at time $t$.  Then a reset occurs during the epoch with probability at most $1- (q/2)$ for some constant $q>0$ depending only on $\dd$.
}
{
Imagine, for the sake of the proof, we flip coins with probability of heads (corresponding to a full slot) $p = 1 - e^{-d}$ for each slot of the epoch in advance.  The sending probability for a packet $u$ in a data slot is $d/s_u$ for some constant $d\leq 1/2$. The probability of sending in each of the \defn{initial} $16/p$ slots is $d \prod_{j=1}^{16/p} (1/j)  = \frac{d}{(16/p)!} = \Theta(1)$ and denote this (small) probability by $q$. By~\lemref{trialsuccesses}, with probability at least $q/2$, the number of heads in any prefix of size $i$ is at least $ip/4$.  Consider the case where this good event occurs (i.e., with probability $q/2$).

Now consider executing the protocol.  Consider some slot $t'$, assuming that there has been no reset in the epoch prior to $t'$.  In slot $t'$, as long as there have been no prior resets, we know that $X(t') \geq X(t)/8$, by \lemref{contentionlower}.  Thus, by \lemref{nonempty}, a slot is empty with probability at most $e^{-d}$, i.e., it is full with probability at least $p$.  Thus, time slot $t'$ is empty only if the  coin flip for slot $t'$ is a tails.  This implies that since there are at least $(t'-t)p/4$ heads in the interval from $t$ to $t'$, there is no reset in slot $t'$. Continuing inductively, we conclude that if the initial coin flips are good, then there is no reset during the epoch.
}

\comment{
\begin{proof}
Imagine, for the sake of the proof, we flip coins with probability of heads $p = 1 - e^{-1}$ for each slot of the epoch in advance.  

The sending probability for a packet $u$ in a data slot is $d/s_u$ for some constant $d<1$. The probability of sending in each of the first $16/p$ slots is $d \prod_{j=1}^{16/p} (1/j)  = \frac{d}{(16/p)!} = \Theta(1)$ and denote this probability by $q$. By~\lemref{trialsuccesses}, with probability at least $p'/2$, the number of heads in any prefix of size $i$ is at least $ip/4$.  Consider the case where this good event occurs (i.e., with probability $7/8$).

Now consider executing the protocol.  Consider some slot $t'$, assuming that there has been no reset in the epoch prior to $t'$.  In slot $t'$, as long as there have been no prior resets, we know that $X(t') \geq X(t)/8$, by \lemref{contentionlower}.  Thus, by \lemref{nonempty}, a slot is empty with probability at most $e^{-d}$, i.e., it is full with probability at least $p$.  
Thus, time slot $t'$ is empty only if the initial coin flip for slot $t'$ is a tails.  This implies that since there are at least $(t'-t)p/4$ heads in the interval from $t$ to $t'$ and hence there is no reset in slot $t'$.

Continuing inductively, we conclude that if the initial coin flips are good (i.e., with probability $1/8$), then there is no reset during the epoch.
\end{proof}
}

\subsection{Successful Streaks}\vspace{\sval}

\seclabel{streaks}

The notation $S(t)$ refers to a streak beginning at time $t$ and continuing for $\sigma_t$ slots.  A streak is \defn{successful} if there are no resets or packet activations during the streak.  Notice that during a successful streak, we know that the contention is always at least $X(t)/8$ (by \lemref{contentionlower}), and that at the end of the streak it is at most $3X(t)/4$ (by \lemref{contentionupper}). We say that  successful streaks  $S(t)$ and $S(t')$ are consecutive if $t' = t + \sigma_t$.\vspace{\sval}

\LemmaProofLater{sigmastreak}{
Let $S(t_1), ..., S(t_k)$ be a set of consecutive streaks for which there is non-zero contention over each streak. Then, for $j=2, ..., k$,   $\sigma_{t_j} \geq  \sum_{i < j} \sigma_{t_i}$.
}
{
Starting from slot $t_1$, consider the set of active packets after $ \sum_{1 \leq i < k} \sigma_{t_i}$  slots for any $k\leq j$. Since the contention is non-zero, there exist remaining active packets, and the  age of each such remaining active packet has increased by $\sum_{1 \leq i < k} \sigma_{t_i}$. There are no injections over successful streaks, therefore, $\sigma_{t_k} \geq  \sum_{1 \leq i < k} \sigma_{t_i}$ for any $k\leq j$.
}

\noindent We now show a key result: an epoch is ``successful'' with constant probability. It follows from an analysis of the change in contention and a union bound over the streaks.

\LemmaProofLater{consecutivestreaks}{
For a non-unit epoch beginning at time $t$, with constant probability, every streak is successful.\vspace{\sval}
}
{
Let $t$ be any slot where $X(t) \geq 8$. Let $S(t_1), ..., S(t_k)$ be consecutive streaks such that $k$ is the first index in these consecutive streaks where  the contention drops below $8$. Define $\tau_j = \sum_{i \leq j } \sigma_{t_i}$. Note that since $X(t) \geq 8$, then $\sigma_t \geq 2$. Then, by~\lemref{noinject}, the probability of a control failure over the interval $[t, t + \tau_k]$ is at most:
\begin{eqnarray*}
& & \left(\frac{1}{2}\right)^{b\cdot X(t)}  + ...  +\left(\frac{1}{2}\right)^{b\cdot X(t+\tau_{k-3})}  +\left(\frac{1}{2}\right)^{b\cdot X(t+\tau_{k-2})}  + \left(\frac{1}{2}\right)^{b\cdot X(t+\tau_{k-1})} + \left(\frac{1}{2}\right)^{b\cdot X(t+\tau_{k})}
\end{eqnarray*}

\noindent  By assumption, $8 \leq X(t+\tau_{k-1})$ and, by \lemref{contentionupper}, we know that $X(t+\tau_{k-1}) \leq \frac{3}{4}\cdot X(t+\tau_{k-2})$. Therefore $\frac{4}{3}\cdot 8 \leq \frac{4}{3}\cdot X(t+\tau_{k-1}) \leq X(t+\tau_{k-2})$. Similarly, we have that $X(t+\tau_{k-2}) \leq \frac{3}{4}\cdot X(t+\tau_{k-3})$. Therefore, $X(t+\tau_{k-3}) \geq \frac{4}{3} \cdot X(t+\tau_{k-2}) \geq (\frac{4}{3})^2 X(t+\tau_{k-1}) \geq  (\frac{4}{3})^2 \cdot 8 $, and generally,  $X(t+\tau_{k-j}) \geq (\frac{4}{3})^{j-1} \cdot 8$. Therefore, we can rewrite the terms as:\vspace{\sval}
\begin{eqnarray*}
& \leq & \left(\frac{1}{2}\right)^{b\cdot X_t }  + ...  + \left(\frac{1}{2}\right)^{(\frac{4}{3})^2\cdot 8 \cdot b} + \left(\frac{1}{2}\right)^{(\frac{4}{3})\cdot 8 \cdot b}  + \left(\frac{1}{2}\right)^{8\cdot b} + \left(\frac{1}{2}\right)^{b}\\
& \leq & \sum\limits_{j=0}^{\log_{4/3} (X_{t})} \left( \frac{1}{2^{ \Omega(b)}}\right)^{(4/3)^j} \\
& \leq & \delta \mbox{~for any arbitrarily small constant $\delta>0$ depending only on sufficiently large $b$ (depending only on $c$)}
\end{eqnarray*}

\noindent Therefore, starting at time slot $t$, the probability that a control failure occurs in the interval  $[t, t + \tau_k]$ defined by these consecutive streaks is at most $\delta$. By \lemref{epochrestart}, the probability of a restart is at most a constant $q/2$ depending only on $d$. Therefore,  the epoch is successful with probability at least $1 - ( (1-\frac{q}{2}) + \delta) = q/2 - \delta \geq \epsilon$ for some constant $\epsilon>0$   depending only on $c$  and $d$. 
}

\noindent We say that an epoch is \emph{disrupted} if at least $1/4$ the slots in the epoch are disrupted.
The following corollary shows that we get constant throughput in an epoch with constant probability. 

\vspace{\sval}

\CorollaryProofLater{epochsgood}{
For a unit epoch that is not disrupted, with constant probability a
packet broadcasts.  For a non-unit epoch with length $T$, with constant probability: (i) every streak in the epoch is successful; (ii) the last streak  is of length at least $T/2$; (iii) the contention throughout the last streak is between $1$ and $256$; and (iv) if the epoch is not disrupted, then at least $\Omega(T)$ packets are broadcast.
}
{
Conclusion (i) follows from \lemref{consecutivestreaks}; conclusion (ii) follows from \lemref{sigmastreak}.  Conclusion (iii) follows because there are no resets or packet activations; hence the contention decreases by at most a factor of 16; however, since the epoch ends, we know that it is no greater than 16 when the last streak ends.  Conclusion (iv) follows from observing that, in an non-disrupted slot, there is a constant probability that a packet is broadcast, and a constant probability that one is not (due to an empty slot or a collision).  If at most $1/4$ of the epoch is disrupted, then at most half of the slots in the last streak are disrupted, and of these $T/4$ non-disrupted slots in the last epoch, in expectation, only a constant fraction are \emph{not} successful broadcasts.  Thus, by Markov's inequality, with constant probability, at most a constant fraction of these $T/4$ slots are not successful broadcasts, and hence with constant probability, at least $\Omega(T)$ packets are broadcast.
}

\vspace{\mval}

\vspace{\sval}

\subsection{Bad Borrower Game}
\seclabel{badborrower}

We have shown that each epoch is \defn{good} 
(satisfying \corref{epochsgood}) with constant probability.
We now abstract away some details, defining a simple game between two players: the \defn{lender} and the \defn{borrower}.  There are two key parameters: a probability $p$ and a fraction $\alpha \in (0,1)$.  The game proceeds in iterations, where in each, the borrower borrows an arbitrary (adversarially chosen) amount of money from the lender, at least one dollar.  With probability $p$, at the end of the iteration, the borrower repays a fraction $\alpha$ of the money. 

The correspondence to our situation is as follows: each iteration is associated with an non-disrupted epoch, the length of the epoch defines the money borrowed, and the number of successful broadcasts defines the money repaid.  In a good epoch, which occurs with constant probability $p$, if the epoch is not disrupted, then we get constant throughput and hence the borrower is repaid an $\alpha$ fraction of her money.  In a bad epoch, by contrast, we allow for the worst case, which is no money paid back at all (no throughput, all \waste).  

For the \defn{finite Bad Borrower game}, there is a predetermined maximum amount that the lender can be repaid: after the borrower has been repaid $n$ dollars, the game ends.  This corresponds to a finite adversary that injects exactly $n$ packets.   In the \defn{infinite Bad Borrower game}, the game continues forever, and an infinite amount of money is lent.  This corresponds to infinite instances, where 
the adversary injects packets forever.

\vspace{-1em}

\subsection{Finite Bad Borrower Game}\vspace{\sval}

Our goal in this section is to show that, when the borrower has repaid $n$ dollars, he has borrowed at most $O(n)$ dollars.  This corresponds to showing that $n$ packets are successfully broadcast in $O(n)$ time, ignoring the interstitial slots (which we will come back to later). We assume throughout this section that $n$ is the maximum amount of money repaid throughout the game, i.e., the adversary injects $n$ packets in an execution.  There is a simple correspondence lemma which bounds the amount of money that the borrower can borrow:\vspace{\sval}
\LemmaProofLater{borrowerbound}{
In every iteration of the finite bad borrower game, $\geq 1$ dollar and $\leq n$ dollars are borrowed.\vspace{\sval}
}
{
The fact that the borrower borrows at least one dollar follows by definition.  Assume the borrower borrows $n$ dollars, i.e., that the associated epoch lasts for at least $n$ slots.  Recall that the last streak in the associated epoch must have been at least $n/2$ slots, and at the beginning of that final streak, the contention must have been at least 8 (or the epoch would have ended).  Since the last streak is of length at least $n/2$, there must be a set of old packets with age $\geq n/2$ that collectively have contention at least $4$ (by definition of a streak).  This implies there must be at least $2n$ such packets, which is impossible, given the bound of $n$ packets total.  Thus, it is impossible to have an epoch of length $n$, and hence to borrow more than $n$ dollars in an iteration of the finite bad borrower game. 
}

\noindent We now argue, via an analysis of the expected repayments, that when the finite bad borrower game ends, the expected cost to the lenders is $O(n)$:\vspace{\sval}

\LemmaProofLater{badborrower}{
Over an execution of the bad borrower game, the expected number of dollars borrowed is~$O(n)$.\vspace{\mval}
}
{
We analyze the dollars repaid in the following fashion: we assume that for every dollar lent, it is paid back with probability $p\alpha$.  Notice, of course, that these random choices are correlated: for a given iteration, either an $\alpha$ fraction of the dollars are paid back (with probability $p$), or no dollars are paid back, with probability $1-p$.  For a given iteration of the game, if there are $k$ dollars borrowed, we see that the expected number of dollars repaid is $pk\alpha$, as expected.

We now ask, what is the expected number of dollars we have to lend in order for $n$ dollars to be repaid?  The answer is $n / (p\alpha)$, i.e., after $O(n)$ dollars have been borrowed, all $n$ dollars have been repaid.  In the last iteration, there can be at most $n$ additional dollars lent (as part of the iteration where the last dollar is repaid), by \lemref{borrowerbound}, yielding an expected number of borrowed dollars of $O(n) + n$.
}

\subsection{Infinite Bad Borrower Game}\vspace{\sval}
\seclabel{infinite}

In order to analyze an infinite executions, we  look at the infinite bad borrower game.  Recall that, 
for parameter $k$ chosen in advance, 
if there have been $k$ dollars borrowed up to some point, then in expectation there have been $O(kp\alpha)$ dollars repaid.  In the infinite case, we conclude something stronger: there are an infinite number of times where the borrower has repaid at least an $\alpha/2$ fraction of the total dollars borrowed.\vspace{\sval}

\LemmaProofLater{infiniteBBgame}{
For all iterations $r$ of the infinite bad borrower game, there is some iteration $r' > r$ such that if the lender has lent $k$ dollars through slot $r'$, then the borrower has repaid at least $kp\alpha/2$ dollars, with probability 1.\vspace{\sval}
}
{
We can look at the random process as a one-dimensional biased random walk with 
variable step size.  Let $X_r$ be the value of the random walk in slot $r$, where $X_0 = 0$.  
Assume we lend $x$
dollars in iteration $r$.  With probability $p$, we succeed in slot $r$ and
hence we define $X_r = X_{r-1} - x + 2x/p$; otherwise, with probability $(1-p)$ 
we define $X_r = X_{r-1} - x$.  

Notice that we have renormalized the random walk, so that every dollar paid 
back is worth $2/(p\alpha)$, i.e., if we have been paid back a
$p\alpha/2$ fraction of the money, then our random walk is at zero.  Thus if we show that
the random walk is positive infinitely often, then we have completed the proof.

(Note that we cannot say that the random walk eventually remains always positive 
from some point on, as
would be true of a simple constant-step-size random walk, because the adversary
can always adjust the step size, for example, employing the following strategy: 
in each step where the random walk is positive, lend twice as much money until 
you lose and the random walk goes negative.)

To analyze this random walk, break the sequence of steps up into
blocks in the following manner.  If the previous block $B_i$ ended with the random walk positive or zero, then block $B_{i+1}$ contains only one step, i.e., the next step of the random walk.  If the previous block $B_i$ ended with the random walk negative, i.e., at $-b_i$, then continue the next block $B_{i+1}$ up until the point where the adversary has cumulatively loaned $2b_i$; notice that in expectation, the 
random walk will increase by $2b_i$, ending the block positive at $b_i$.  (Since the execution is infinite, and since the lender  has to lend at least one dollar in each step, eventually every block will end.) 

We now use a Hoeffding's inequality to show that with constant probability, the random walk returns to zero at the end of every block.  First, if block $B_{i+1}$ begins with the random walk zero or positive and takes only one step, then with constant probability that step is positive.  Next, consider the case where $B_{i+1}$ begins with the random walk negative, and lends at least $2b_i$ cumulatively throughout. 

Let $Z_1, Z_2, \ldots, Z_k$ be the random variables associated with the change in value at each step of the random walk in the block, where in step $Z_j$ the lender lends $z_j$ dollars; with probability $p$, $Z_j = 2z_j/p - z_j$, and with probability $(1-p)$, $Z_j = -z_j$.  Thus each $Z_j$ has a bounded range $|Z_j|$ of size $2z_j/p$.  

Let $\bar{z} = \sum_j z_j$, and recall that by construction, $\bar{z} \geq 2b_i$.  Since $|Z_j| = 2z_j/p$, we know that $\sum |Z_j|^2 \leq 4\bar{z}^2/p^2$.  Finally, we observe that for each $Z_j$, the expected value is $z_j$, and the expected value of the sum is~$\bar{z}$.  

Since the success or failure of epochs is independent (as packets are making independent choices in each slot), we can apply Hoeffding's inequality to lower bound the sum, where we choose $t = \bar{z}/2$:
\begin{eqnarray*}
\Pr((\sum_j Z_j - \bar{z}) \leq -t) & \leq & e^{-\frac{2t^2}{\sum_j|Z_j|^2}} \\
& \leq & e^{-2p^2t^2/(4\bar{z}^2)} \\
& \leq & e^{-p^2/8}
\end{eqnarray*}
That is, with constant probability, the random walk gains at least $\bar{z}/2 \geq b_i$ during this block, and hence returns to zero.

To conclude the proof, we observe that since each block ends with the random walk returning to zero, there are an infinite number of points where the random walk returns to zero.  (We cannot bound the length of time it takes to return to zero without first bounding the amount of money that can be lent.)  Assume that at the end of block $B_i$, the random walk has returned to zero, and over the entire execution up until that point, the lender has lent $k$ dollars.  Since each dollar paid back causes the random walk to increase by $2/(p\alpha)$, this means that $kp\alpha/2$ dollars have been repaid, as required.
}

\noindent As a corollary, if we only consider the epochs, ignoring the contribution from
the interstitial slots, we can show constant throughput for infinite executions. 
Specifically, we can use 
the infinite bad borrower game to define ``measurement points,''
thus showing that in an infinite execution, there are an infinite number
of points at which we get constant throughput (if we ignore the 
contribution from the interstitial slots).  \vspace{\sval}

\begin{corollary}
\corlabel{infiniteEpochThroughput}
If we take an infinite execution and remove all slots that are not part of an epoch,
then the resulting execution has at most a constant fraction of \waste.
\vspace{\sval}
\end{corollary}

\noindent We later show (\subsecref{interstitial-infinite})
that the contribution from the
interstitial slots does not hurt the \waste, meaning that 
we get at most a constant factor of \waste taking into account all slots.

\vspace{\mval}

\subsection{Interstitial Slots, Expected \Waste, and Expected Throughput for Finite Instances}\vspace{\sval}

\subseclabel{interstitial-finite}

We begin by considering the finite case where there are $n$ packets injected. We bound the length of the interstitial slots, after which we prove at most an 
expected constant factor of \waste.  

We first argue that for a packet's lifetime, any prefix of at least
2 slots is at least a constant fraction full. 

\vspace{\sval}

\LemmaProofLater{packetprefix}{
Suppose that a packet $u$ is active for the time interval $[t,s]$.
Then for any time $t'$ with $t < t' \leq s$, at least a $1-\gamma =
1/16$ fraction of the slots in the interval $[t,t']$ are full. }{
The proof is by contradiction.  Suppose that $[t,t']$ is strictly less
than a $(1-\gamma)$-fraction full.  Let $x = t'+1-t$ be the total
number of slots in the interval, and let $k$ be the number of full
slots in the interval.  Then we have $x > \frac{1}{1-\gamma}k = 16k$.
Since $x>16k$ is an integer, $x \geq 16k+1$. Thus, the subinterval
$[t,t'-1]$ contains at most $k$ full slots across $x-1\geq 16k$ slots,
meaning that a reset would occur at or before time $t'-1$. 
}

Our next lemma extends the above argument to cover all interstitial
slots.  We would like to say that the first $t$ timeslots of the
entire execution include at most a $\gamma$-fraction of empty slots.  This is not
necessarily true --- the first slot could be empty.  The
issue is that \lemref{packetprefix} does not apply to the first step
of a packet's lifetime.  But we can make a similar claim if we elide
certain slots.  We define a slot to be \defn{active}
if at least one packet is active, and we define a slot to be a \defn{quiet
  arrival} if a packet activates but the slot is empty.  The following
lemma achieves our goal by ignoring slots with quiet arrivals.

\LemmaProofLater{boundempty}{ For any integer $t>0$, consider the first $t$
  active time slots that are not quiet arrivals.  At most a
  $\gamma =15/16$-fraction of these slots is empty.  } {
  The proof is by induction on $t$.  For the base case, observe that a
  quiet arrival results in an immediate reset of that packet.  Thus,
  the first time step in consideration is a step during which a packet
  arrives and the slot is full.  

  For the inductive step, we assume that the claim holds for all of
  the $t$ first steps and argue that it holds at $(t+1)$-th. Consider
  any packet $u$ that is active at time $t+1$.  Let $t_u \leq t+1$ be
  the step at which $u$'s current lifetime began.  We have two cases.\\
  \noindent Case 1. If $t_u = t+1$, then the packet just activated.
  By assumption, this is not a quiet arrival, and hence the $(t+1)$th
  step is full.  By inductive assumption, the first $t$ slots are at
  most a $\gamma$-fraction empty.  Concatenating these slots proves
  the claim.\\
  \noindent Case 2. If $t_u < t+1$, we are thus considering a
  length-$\geq 2$ prefix of the packet's lifetime, except that any
  quiet arrivals (i.e., empty slots) therein have been elided.
  Applying \lemref{packetprefix}, we conclude that at most a $\gamma$
  fraction of these slots are empty.  By inductive assumption, at most
  a $\gamma$ fraction of the slots up to $t_u-1$ are also empty.
  Concatenating these slots proves the claim.
}

\noindent Observe that \lemref{boundempty} counts all of the active
interstitial slots, as any quiet arrivals are by definition part of an
epoch.  We thus have a way of charging the empty interstitial slots
against full slots, incurring at most a
$1/(1-\gamma)$ cost.

Our goal now is to bound the number of full interstitial slots,
specifically the non-disrupted slots.  The main idea is to show that
for each non-empty and non-disrupted slot, there is a constant
probability of successful transmission.  Thus after
$O(n)$ such slots, in expectation, all the packets have
broadcast.\vspace{\sval}

\LemmaProofLater{collisionSuccess}{
For a slot $s$, let $e_{\geq 2}$ denote the event where two or more packets are broadcast in $s$, and let $e_{=1}$ denote the event where one packet is broadcast in slot $s$. If $s$ is an interstitial slot, then $Pr(e_{\geq 2}) = O(Pr(e_{=1}) )$.\vspace{\sval}
}
{
The lemma follows immediately from \lemreftwo{efficiency}{nonempty}, since in interstitial slots, the 
contention is $O(1)$.
}

\LemmaProofLater{fullFiniteInterstitial}{
There are at most $O(n)$ full, non-disrupted interstitial,
slots in expectation. \vspace{\sval}
}
{
By the time that there are 
$n$ full slots that have successful transmissions, the execution is over.
And if we condition upon a given slot being full, there is a constant
probability of a successful transmission by~\lemref{collisionSuccess}.  Thus it takes $O(n)$ such slots, in expectation, before all $n$ packets have successfully transmitted.
}

\noindent We can now prove our claims in~\thmref{finite} and
\corref{finite} regarding expected \waste and throughput:\vspace{\sval}

\LemmaProofLater{finite}{
If the adversary injects $n$ packets,~\Alg~has at most an expected constant factor \waste.\vspace{\sval}
}
{
We will argue that the expected number of slots for all the packets to
finish is: $\expect{T} = O(n + \Cost)$ slots. We then observe that the
expected \nonwaste is $\expect{\lambda} = \expect{(n+\Cost)/{T}}$,
which by Jensen's inequality is at least a constant.  Throughout the
proof we consider only active slots.  Reincorporating the inactive
slots only increases the waste by a constant factor as a packet
activates after seeing an inactive slot. 

Let $n_e$ denote the total number of slots over all non-disrupted
epochs\footnote{Recall that an epoch is non-disrupted if $< 1/4$ of
  its slots are disrupted.}, let $n_d$ denote the number of slots over
all disrupted epochs, and let $n_i$ denote the number of full
non-disrupted interstitial slots.  Again, let $\Cost$ denote the total
number of disrupted slots.

Our goal is to bound the number of empty interstitial slots.
\lemref{boundempty} implies that, ignoring some empty epoch slots
(namely, the quiet arrivals), at most a $\delta$-fraction of the
remaining slots are empty.  In particular, the worst case occurs if we
pessimistically assume all epoch slots are full, giving at most
$O(n_e+n_d+n_i+\Cost)$ empty interstitial slots.

By \lemref{badborrower}, the finite bad borrower game implies that the
number of epoch slots in non-disrupted epochs required to complete all
$n$ packets is $O(n)$ in expectation, therefore, $E[n_e] = O(n)$. As
for the interstitial slots,~\lemref{fullFiniteInterstitial} shows that
$E[n_i] = O(n)$.  Therefore, among non-disrupted epochs and
non-disrupted interstitial slots, we conclude that the expected number
of slots required for all $n$ packets to succeed is $O(n)$.

Finally, we count the number of disrupted slots.  Since a disrupted
epoch is one in which at least $1/4$ of the slots are disrupted, there
are clearly at most $O(\Cost)$ disrupted epoch slots.  Similarly,
there are at most $O(\Cost)$ disrupted, non-empty insterstitial slots.
There are also at most $O(\Cost)$ slots in which the control channel
is disrupted (which can cause wasted time on the data channel if there
are no active packets).  Thus, there are at most $O(\Cost)$ such slots
otherwise unaccounted for.

Thus we conclude that there are, in expectation, $O(n)$ non-disrupted epochs and non-disrupted interstitial slots, at most $O(\Cost)$ disrupted slots and disrupted epochs, and $O(n+\Cost)$ empty slots. 
}

\vspace{\mval}

\subsection{Interstitial Slots for Infinite Instances}\vspace{\sval}
\subseclabel{interstitial-infinite}

We now show that the contribution from the interstitial
slots does not hurt the throughput in infinite executions.  
To do so, we deterministically bound
the contribution from the empty interstitial slots.  We
show that as long as we pick ``measurement points'' that are sufficiently
large that from then on the non-empty interstitial slots do not hurt. 
We use the following well known facts about random walks~\cite{mitzenmacher:probability}. \vspace{\sval}

\begin{fact}
Suppose that we have a biased random walk on a line with fixed step size, where the
probability of going right is at least $p$, the probability of going left is at most
$1-p$, and the step size right is $\delta_r$ and the step size left is
$\delta_\ell$.  Suppose that $p\delta_r> (1-p)\delta_\ell$.  Then if the
random walk starts at the origin, the probability of returning to the
original is some constant strictly less than $1$. \vspace{\sval}
\factlabel{random-walk-fact} \end{fact}

\begin{corollary}
For any such biased random walk, there there is a last time that the walk returns to the origin.\corlabel{random-walk-cor} 
\end{corollary}

\noindent We use \corref{random-walk-cor} to bound the ratio of
collisions to broadcasts in non-disrupted, non-empty interstitial
rounds.  From some point on, the number of collisions is
always at most a constant factor of the number of broadcasts, and
hence yields constant throughput.  We again observe that the empty
slots cannot hurt the throughput by more than a constant factor
overall, because, by \lemref{boundempty} at most a $\gamma$ fraction
of the slots in any prefix can be empty interstitial slots.  Finally,
we bound the disrupted interstitial slots by $\Cost$.  From this,
we conclude that we achieve constant \nonwaste and throughput, as
claimed in \thmref{infinite} and \corref{infinite}:\vspace{\sval}

\LemmaProofLater{infinitethroughput}{
In an infinite execution,~\Alg~achieves at most a constant-fraction of \waste.\vspace{\sval}
}
{
For some slot $t$, let $i^b_t$ be the number of successful broadcasts in non-disrupted interstitial slots prior to time $t$, and let $i^c_t$ be the number of collisions in non-disrupted interstitial slots prior to time $t$.

We first argue that, with probability 1, from some point $t$ onwards, for all $t' > t$: $i^c_t \leq O(i^b_t)$.  Conditioned on the fact that there is at least one broadcast in an non-disrupted interstitial round, let $p$ be the probability of a successful broadcast and $q = 1-p$ be the probability of a collision.  We know from \lemref{collisionSuccess} that $q = O(p)$.

Define the following random walk: with probability $q$ take a step to the left of size 1, and with probability $p$ take a step to the right of size $2q/p$.  Since $p\cdot(2q/p) > q$, by \corref{random-walk-cor} we know that from some point on, this random walk is always positive.  

Let $t$ be a time slot that is after the last point where the random walk crosses the origin.  We can then conclude that $i^c_t < i^b_t \cdot (2q/p)$.  Since $2q/p = O(1)$, we conclude that $i^c_t = O(i^b_t)$.

Finally, we analyze the throughput.  Fix any time $t$.  Let $\hat{t}$
be the smallest time after $t$ where the random walk defined above is
positive.  According to \lemref{infiniteBBgame}, there is a time
$t' > \hat{t}$ where we have achieved constant throughput during the
non-disrupted epochs, i.e., a constant fraction of the slots in
non-disrupted epoch are broadcasts.  By the analysis of the random
walk, we conclude that a constant fraction of the non-empty,
non-disrupted interstitial rounds are broadcasts.  By assumption, at
most $O(\Cost)$ slots are part of disrupted epochs, and there are at
most $\Cost$ disrupted interstitial slots.  There are at most $\Cost$
disrupted control slots (which may cause delays on the data channel if
there are no active packets).  Finally, by \lemref{boundempty}, at
most $\gamma$ of the data channel slots in the entire execution are
empty interstitial slots.  Putting these pieces together yields
constant throughput overall.  }
\vspace{\sval}

\vspace{\mval}

\section{Analysis of the Number of Access Attempts}\vspace{\sval}
\seclabel{energy}

In this section, we analyze the number of access attempts.  Our goal is to show that in the absence of disruption, the number of broadcasts is small, and that the adversary requires a significant amount of disruption to cause even a small increase in the number of access attempts.

We first analyze how often a packet resets, showing that it is likely to succeed before it has a chance to reset.  This ensures that a packet cannot be forced to make a large number of attempts via repeated resets.\vspace{\sval}
\LemmaProofLater{successBeforeReset}{
  With constant probability, a packet succeeds before it
  resets.\vspace{\sval}
}
{
  In this proof, we grant the adversary even more power than given by
  the model---in each slot, the adversary is allowed to specify
  whether the slot is ``covered'', meaning that it is either disrupted or
  some other packet transmits.  The only thing the adversary does not
  control is the packet in question.
    
  Starting from the time the packet becomes active, we divide time
  into windows $W_0,W_1,W_2,W_3,\ldots$, where window $W_i$ has
  length~$2^i$.  Note that if the first slot is covered, the packet
  cannot possibly reset until time 8 or later, which more than subsumes window $W_1$.
  Similarly, if at least 1 slot is also covered in window $W_1$, then
  the packet cannot possibly reset until after window $W_2$.  In
  general, if at least half the slots are covered in each of the
  windows $W_0,W_1,\ldots,W_{i-1}$, then the packet either stays alive
  through $W_i$, or it succeeds sometime before the end of $W_i$---it
  cannot reset.  Our argument thus proceeds inductively over windows,
  stopping at the first window $W_i$ that is not at least half covered.  In
  window $W_i$, the packet transmits independently in each data slot
  with probability at least $d/2^{i+2}$, where $d$ is a constant
  specified in the protocol.  Thus, if at least $2^{i-1}$ of the slots
  in $W_i$ are left uncovered, the packet has either succeeded
  earlier, or it succeeds in window $W_i$ with probability at least
  $1-(1-d/2^{i+2})^{2^{i-1}} \geq 1-1/e^{d/8}$, which is constant. 
}

\CorollaryProofLater{fewresets}{
  For any positive integer $k$, the probability that a packet resets
  $k$ times is at most $1/e^{\Theta(k)}$. \vspace{\sval}
}
{
  The only observation we need is that for a particular packet, each
  of its lifetimes are nonoverlapping. Thus, each
  trial of \lemref{successBeforeReset} is
  independent.  Each reset occurs with probability at most
  $1/e^{d/8}$, and hence the probability of $k$ resets is at most
  $1/e^{kd/8} = 1/e^{\Theta(k)}$.\vspace{\sval}
}

\noindent We can now bound the total number of access attempts that a packet makes during the first $t$ slots after its arrival.  If it has not yet reset by time $t$, it is easy to see that it has made $O(\log^2{t})$ access attempts in expectation---and we have show above that a packet is unlikely to reset too many times.  This yields:\vspace{\sval}
\LemmaProofLater{packetEnergy}{
  In the first $t$ slots following a packet's arrival, it makes
   $O(\log^2(t))$ attempts in expectation.\vspace{\sval}
}
{
Consider any lifetime of the packet.  The expected number of access attempts during a
  slot is equal to the packet's transmission probability, and hence
  the expected total number of access attempts is the sum of probabilities across all
  slots by linearity of expectation.  The expected number of access attempts in a
  lifetime is thus at most $\sum_{s=1}^t \Theta(\ln s)/s =
  \Theta(\log^2 t)$, with the $\ln s$ arising from the high
  transmission probability in control slots.

  We now compute the expected number of access attempts made by the packet by using linearity
  of expectation across all lifetimes of the packet. In particular,
  the number of access attempts during the $k$th lifetime is 0 if the packet does not
  reset $k-1$ times, and hence applying \corref{fewresets} the
  expected number of access attempts of the $k$th lifetime is $1/e^{\Theta(k)} \cdot
  O(\log^2t)$.  Using linearity of expectation across all lifetimes,
  we get a total expected number of access attempts of $O(\log^2 t) \sum_{k=0}^{\infty}
  1/e^{\Theta(k)} = O(\log^2 t)$.  
}

\noindent We can now prove our claims in~\thmreftwo{finite}{infinite} regarding the expected number of access attempts per packet.  In the finite case, we have already bounded the expected length of the execution in \lemref{finite}.  In the infinite case, we separately analyze the disrupted slots, the non-disrupted slots with young packets, and the non-disrupted slots with old packets, bounding the number of attempts.\vspace{\sval}

\LemmaProofLater{finiteEnergy}{
  Consider the finite case, let $n$ be the number of injected packets,
  and let $\Cost$ be the number of disrupted slots.  Then the expected number of access attempts
  each packet makes is $O(\log^2(n+\Cost))$.  \vspace{\sval}
}
{
  Suppose the execution completes in time~$t$.  Then applying \lemref{packetEnergy}
  yields an expected number of access attempts of $O(\log^2 t)$, as the packet must
  complete before the execution completes.  Let $T$ be the expected time of completion.  From
  Markov's inequality, $t\geq T^i$ with probability at most
  $1/T^{i-1}$.  Summing across all $i$, the expected number of access attempts becomes
  $O(\sum_{i=1}^\infty (1/T^i)\log^2(T^i)) = O(\log^2
  T)\cdot\sum_{i=1}^\infty (i/T^i) = O(\log^2 T)$.  Substituting in
  the expected makespan $T = O(n+\Cost)$ from~\lemref{finite} concludes the theorem.
}

\LemmaProofLater{infiniteEnergy}{
  Consider any time $t$ in the infinite case at which we have $\lambda$
  throughput, for constant $\lambda$.  Let $\Cost_t$ be the total number
  of disrupted slots before $t$, and let $\nc_t$ be the maximum
  contention prior to time~$t$.  Then the expected average number of access attempts per packet
  is $O(\log^2(\nc_t) + \log^2(\Cost_t))$.\vspace{\sval}
}
{
  The analysis is split into two cases: either $\Cost_t\geq \lambda t/2$,
  or $\Cost_t < \lambda t / 2$.  The first case is
  easy---\lemref{packetEnergy} states that the expected number of access attempts per
  packet is $O(\log^2t) = O(\log^2\Cost_t)$.

  Suppose for the remainder that $\Cost < \lambda t/2$.  Then we divide the
   analysis here into three parts: (A) the disrupted slots, (B) the
  non-disrupted slots for ``young'' packets, and (C) the non-disrupted slots for
  ``old'' packets.  In parts~A and~B, we shall show that the expected number of access attempts 
  per packet is at most $O(\log^2\Cost_t + \log^2\nc_t)$, regardless of
  whether we have constant throughput.  It is only in part~C that we
  leverage the assumption that time $t$ is a time at which we have
  constant throughput. 

  A. Consider a single lifetime of a specific packet.  Our goal is to
  bound the number of access attempts by this packet during all $\Cost_t$ disrupted
  slots.  The number of access attempts is maximized if all $\Cost_t$ slots occur as early as
  possible in the packet's lifetime, in which case the expected number of access attempts
  is at most $O(\log^2 \Cost_t)$ per lifetime.  As in
  \lemref{packetEnergy}, we then apply \corref{fewresets} and
  linearity of expectation to get an expected number of access attempts of at most
  $O(\log^2 \Cost_t) \sum_{k=0}^\infty 1/e^{\Theta(k)} = O(\log^2 \Cost_t)$.
  
  B. Consider a specific packet.  We say that the packet is young
  during the first $\nc_t^2$ steps of its lifetime, during which it
  makes $O(\log^2 \nc_t^2) = O(\log^2 \nc_t)$ access attempts in expectation
  (\lemref{packetEnergy}).  Summing across all lifetimes as above, we
  conclude that the contribution for young packets is $O(\log^2
  \nc_t)$.

  C. If a packet is not young, i.e., if its age is at least $\nc_t^2$,
  we say that it is old.  Unlike parts~A and~B which analyze on a
  per-packet basis, this part analyzes the number of access attempts in aggregate.  For
  every non-disrupted slot, there are at most $\nc_t$ packets in the
  system, and hence at most $\nc_t$ packets are old.  Each old packet
  transmits with probability at most $O(\ln(\nc_t^2)/\nc_t^2) =
  O(1/\nc_t)$, and hence the expected number of access attempts across all old packets in
  any slot is $O(1)$.  Using linearity of expectation across all $t$
  slots, we have a total expected number of access attempts by old packets of $O(t)$.  
  
  Since we have constant throughput at time $t$, we know that at least
  $\lambda t$ slots are either disrupted or successful transmissions.
  There are at most $\lambda t/2$ disrupted slots by assumption, and
  hence there are at least $\lambda t/2 = \Omega(t)$ successful
  transmissions.  We charge the $O(t)$ number of access attempts from old packets to
  these $\Omega(t)$ completed packets, for an additional $O(1)$ number of access attempts
  per successful packet.  
}

\vspace{\bval}

\section{Synchronization: Reducing Two Channels to One}
\seclabel{synch-short}

We now briefly describe how to transform the \Alg algorithm so that it runs on a single channel.  Details are provided in \appref{synch}.

As in \Alg, packets are initially inactive.  In this case, they monitor the channel and wait to hear two empty slots, immediately after which they become active. 
Once active, packets alternate executing \emph{control slots} and \emph{data slots}.  A packet calculates its age as follows: prior to its first active slot (which is a control slot), it sets its age to 1; immediately before every subsequent control slot, it increments its age. 
When a packet first becomes active, it treats its first active slot as a control slot.  Moreover, in that one slot, it sends a control signal \emph{with probability $1$}.  It then proceeds to alternate data and control slots.  

With no further synchronization, different packets may treat a given slot as a control slot and a data slot at the same time.  We thus add a synchronization mechanism: \emph{if a packet observe an empty control slot followed by a non-empty data slot, then it performs an additional data slot.}

If a packet completes in a data slot that immediately follows an empty control slot, then it does not terminate immediately, but participates in the second data slot that follows before terminating.

Finally, recall that a packet resets if it finds that a $\gamma$-fraction of data slots have been empty since it became active.  Here, the reset rule remain identical with one change: if there are two consecutive data slots, then the packet does not count the results from the first data slot. 

We observe that packets agree on whether a slot is a control or a data slot, i.e., synchronization works:
\vspace{\sval}
\LemmaProofLater{slotagreement}{
Let $t$ be a slot, and let $u$ and $v$ be two packets that are active in slot $t-1$ and slot $t$.  Then $u$ considers $t$ a control slot if and only if $v$ considers $t$ a control slot.  Similarly, $u$ considers $t$ a data slot if and only if $v$ considers $t$ a data slot.
}
{
Assume $u$ and $v$ are injected in the same slot $t_0 < t$.  Then $u$ and $v$ both consider $t_0$ to be a control slot, and observe the same pattern of full and empty slots from then on.  Hence $u$ and $v$ continue to identify slots in the same manner.

Assume instead that $u$ is injected in slot $t_u$ and $v$ is injected in slot $t_v > t_u$ (where $t_v < t$).  There are two cases.  First, the slot $t_v$ may be considered a control slot by $u$.  In that case, as of slot $t_v$, both $u$ and $v$ consider $t_v$ to be a control slot and both observe the same pattern of full and empty slots from then on.  Hence $u$ and $v$ continue to identify slots in the same manner.

Finally, assume that $u$ considers slot $t_v$ to be a data slot.  We know that $v$ broadcasts in slot $t_v$, because a packet broadcasts in its first control slot, and hence $t_v$ is non-empty.  Notice, though, that $t_v-1$ and $t_v-2$ must be empty slots, as packet $v$ only becomes active after observing two empty slots.  From this we conclude that $u$ considers $t_v-1$ to be a control slot: we know (by assumption) that $t_v$ is a data slot; if $t_v-1$ were also a data slot, then the preceding control slot $t_v-2$ must also have been non-empty (in order to force the repeated data slot); however, we know that $t_v-2$ is empty.  

Since $u$ considers $t_v-1$ is an \emph{empty} control slot, and $u$ considers $t_v$ to be a non-empty data slot, it designated slots $t_v+1$ to be a data slot.  Packet $v$ also considers $t_v+1$ to be a data slot (because it is alternating slot types).  Hence, as of slot $t_v+1$, both $u$ and $v$ agree on the slot designation.  From that point on, but packet $u$ and $v$ observe the same pattern of full and empty slots, and hence they continue to identify slots in the same manner.
}

We can then repeat the analysis found in \secreftwo{analysis}{energy}, with a very small number of minor changes (see appendix), yielding the same waste, throughput, and energy results.


\newpage

\appendix


\section{Bad-Throughput Example for Exponential Backoff}


\begin{figure}[htb]
  \centering
  \includegraphics[width=.7\linewidth]{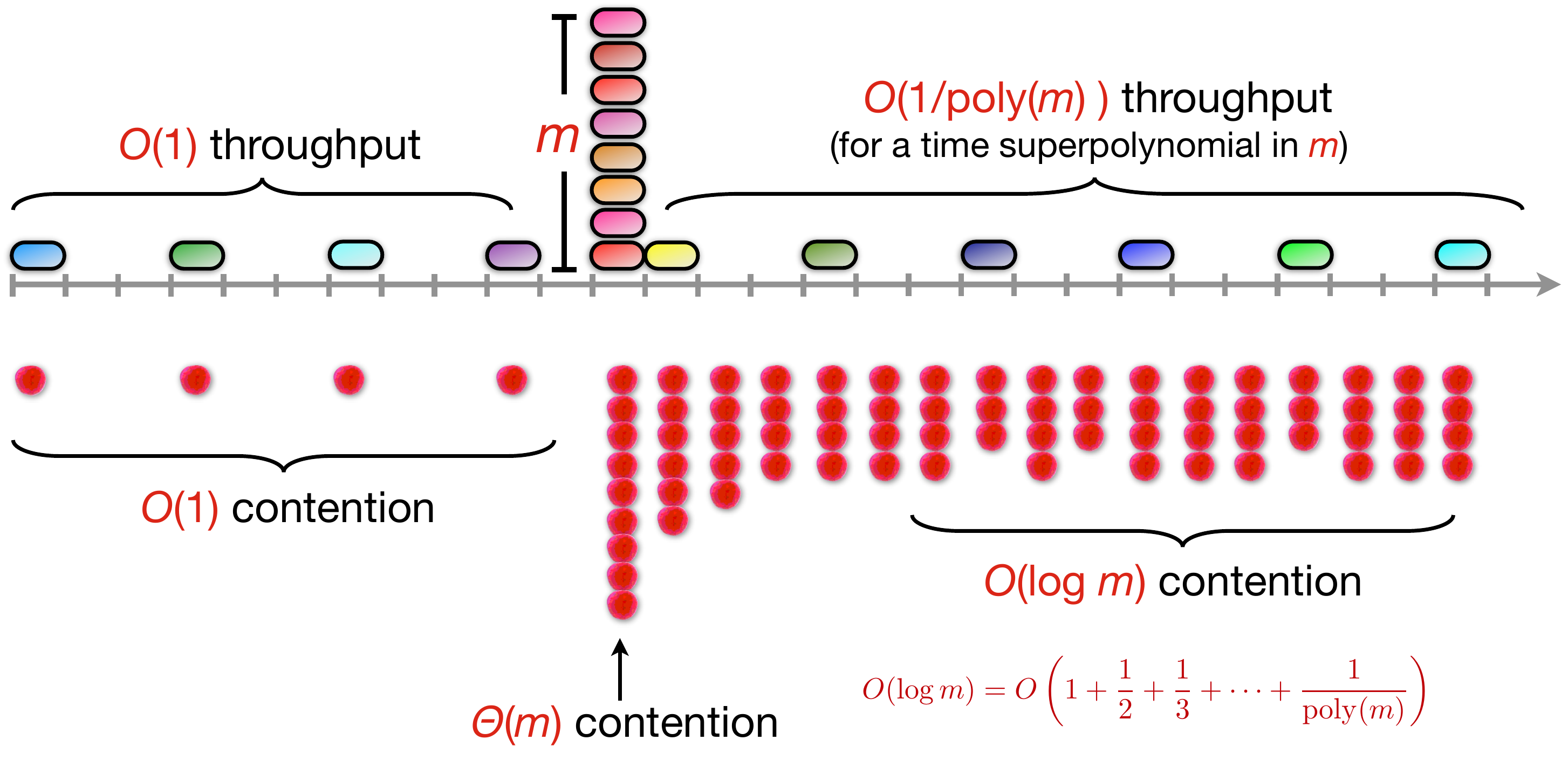}
  \caption{Illustration of how exponential backoff struggles with
    bursts~\cite{BenderFaHe05}. Shown above the line are packet
    arrivals, and below the line are transmission attempts. (When
    there is more than red dot below the line, there is a collision.)
    The packet arrivals illustrated include a steady stream of one new
    packet every three timesteps, plus a single burst of $m$ packets.
    Initially, the throughput is great.  After the burst, the
    contention grows large, and the steady stream is enough to
    maintain the contention at $\Theta(\log(m))$ and reduce the
    throughput to $O(1/\mbox{poly}(m))$ for a
    length of time that is super polynomial in $m$.}
\label{fig:fig-exp-backoff-struggles}
\end{figure}



\newpage


\appendix

\section{Omitted Proofs}
\applabel{proofs}


\subsection*{Preliminaries}
\sk

\subsection*{Individual Slot Bounds}
\efficiency

\nonempty

\subsection*{Contention Bounds}
\contentionupper

\contentionlower

\subsection*{Control Failures}
\singleslot

\noinject

\subsection*{Bounding Resets}
\trialsuccesses

\epochrestart

\subsection*{Streaks}
\sigmastreak

\consecutivestreaks

\epochsgood

\borrowerbound

\badborrower

\infiniteBBgame

\subsection*{Interstitial Slots}


\packetprefix

\boundempty

\collisionSuccess

\fullFiniteInterstitial

\finite

\infinitethroughput


\subsection*{Number of Access Attempts}

\successBeforeReset

\fewresets

\packetEnergy

\finiteEnergy

\infiniteEnergy


\newpage


\section{Synchronization}
\applabel{synch}

In this section, we describe how to transform the \Alg algorithm so that it runs on a single channel.  We first describe the modified algorithm.  We then show that it maintains a synchronized view of the channel.  Finally, we review the analysis and see how it has to be modified for this variant.

\subsection*{\Alg, Take 2}

Here we describe how to modify \Alg to work with only one channel.  

As in \Alg, packets are initially inactive.  They monitor they channel and wait to hear two empty slots.  As soon as they hear two empty slots, they become active in the following time slot.  (Note this differs from \Alg, where a packet becomes active in the next slot when it hears one empty control slot.)

Once active, packets will alternate executing \emph{control slots} and \emph{data slots}.  Recall that these are implemented as follows:
\begin{itemize}
\item \emph{Control slot:} Send control signal with probability $c\max\{\ln{s_u}, 1\}/s_u$.
\item \emph{Data slot:} Send message with probability $d/s_u$.  If it is successful, terminate.
\end{itemize}
A packet calculates its age as follows: prior to its first active slot (which is a control slot), it sets its age to 1; immediately before every subsequent control slot, it increments its age. That is, the age of a packet in slot $s$ is the number of control slots it has seen since it became active, up to and including slot $s$. 

When a packet first becomes active, it treats its first active slot as a control slot.  Moreover, in that one slot, it sends a control signal \emph{with probability $1$}.  It then proceeds to alternate data and control slots.  For example, if packet $u$ is inactive and it observes slot $s$ and $s+1$ to be empty, it becomes active in slot $s+2$.  It treats $s+2$ as a control slot and sends a control signal.  It then alternates, treating $s+3$ as a data slot, $s+4$ as a control slot, etc.

With no further synchronization, different packets may treat a given slot as a control slot and a data slot at the same time.  We thus add a synchronization mechanism:
\begin{itemize}
\item If a packet observe an empty control slot followed by a non-empty data slot, then it performs an additional data slot.
\end{itemize} 
For example, assume packet $u$ believes that $s$ is a control slot.  If $s$ is empty (i.e., no broadcast, no collision, no disruption), and if $s+1$ is non-empty (i.e., a broadcast or a collision or disruption), then packet $u$ treats slot $s+2$ as a data slot.  It then continues to alternate, treating $s+3$ as a control slot.

If a packet completes in a data slot that immediately follows an empty control slot, then it does not terminate immediately, but participates in the second data slot that follows before terminating.

Finally, recall that a packet resets if it finds that a $7/8$-fraction of data slots have been empty since it became active.  Here, the reset rule remain identical with one change: if there are two consecutive data slots, then the packet does not count the results from the first data slot.  For example, if $s$ is a control slot and $s+1$ and $s+2$ are data slots (because $s$ is empty and $s+1$ is non-empty), then after these three slots: if $s+2$ is empty, a packet increments its count of empty data slots by one; if $s+2$ is full, a packet increments its count of full data slots by one.

\subsection*{Synchronization}

We now argue that every packet that has been active for at least one complete slot agrees on whether a slot is a control or a data slot:

\slotagreement

As a result of \lemref{slotagreement}, we can officialy designate slots as control or data slots.  Consider a slot $t$:
\begin{itemize}
\item If no packet is active in slot $t$, we designate slot $t$ to be an empty slot.
\item If there exists any packet $u$ that is active in slot $t-1$ and slot $t$, then we designate slot $t$ a data or control slot based on the designation of packet $u$.
\item If all the packets active in slot $t$ were \emph{not} active in slot $t-1$, then we designate slot $t$ a control slot (as do all the newly activated packets).
\end{itemize}
We can then designate a pair or triple of slots consisting of a control slot followed by one or two data slots as a slot-group.

Throughout, unless specified otherwise, when we refer to a slot as a control or a data slot, we refer to this global (synchronized) designation.

\subsection*{Throughput Analysis}

We now reprise the analysis of \Alg, examining what has changed and what has not.  

\paragraph{Preliminaries.}
We define the contention $X(t)$ in slot $t$ as before, i.e., the sum of the ages of active packets.  Similarly, we define $\sigma_t$ and divide packets into young and old packets in the same manner.  \lemref{sk} holds unchanged.

We now define a control failure to occur in a slot-group if no packet broadcasts during the control slot of that slot-group.  Note that a control failure is not a sufficient condition for a packet to become active, as two consecutive empty slots are needed.  However, it remains a necessary condition because there cannot be two consecutive data slots: a slot-group only contains two data slots if the first data slot is non-empty.

\paragraph{Individual slot calculations.}
In \lemref{efficiency}, we examine the probability that a packet broadcasts successfully; the probabilistic calculation (which depends only on packet ages) remains valid for any data slot in which there is no disruption \emph{and} in which no packet becomes active.  (Note that if a packet becomes active in data slot $t$, then the new packet treats it as a control slot and transmits a control signal, disrupting any and all communication.)  In a slot-group with three slots, when applying \lemref{efficiency}, we will be using it to refer to the second data slot.

Similarly, \lemref{nonempty} is a probabilistic calcuation that depends only on packet ages, and remains valid---as long as no no packet becomes active in that slot.  (In the latter case, the probability of a collision is equal to the probability of some broadcast.)

\paragraph{Epochs, streaks, and interstitial slots.}
As before, we define an epoch to begin when a new packet is activated.  Specifically, we say that an epoch begins at the beginning of a slot group in which some packet is activated.  

If an epoch begins in slot $t$, we define a streak to consist of the subsequent $\sigma_t$ slot-groups (which will consist of somewhere betweent $2\sigma_t$ and $3\sigma_t$ slots, depending on how many slot-groups there are with three slots).  If a streak ends with contention below 32, then the epoch ends and interstital slots begin; otherwise a new streak begins.  (Notice that we have increased the contention limit for an epoch.)

We say say that an epoch is disrupted if at least $1/4$of its data slots are disrupted; for this purpose, if a slot-group contains three slots, we count only the second data slot.

\lemref{contentionupper} shows that if no control failure occurs during a streak, then the contention during the streak drops by at least $1/4$.  This remains true, as it is only a function of young packets aging during the streak and depend only on the definition of $\sigma_t$.

\lemref{contentionlower}, by contrast, shows that the contention does not drop by too much during a streak, as long as there are no resets.  The key difference here is that, due to the extra data slots, more packets can complete, lowering the contention twice as fast.  Thus, for the revised version of this lemma, we assume that initially $X(t) \geq 16$, and conclude that it does not drop by more than a factor of 16.  (In fact, a streak begins with contention at least 32, for reasons we will see later.)

\paragraph{Control failures.}
Next, we look at the probability of a control failure.  In \lemref{singleslot}, we show that if there are no control failures and resets in a streak up to some time $t$, then we can bound the probability of a control failure in the next control slot.  This is simply a probability calculation that depends only on the ages of the packets.  The only difference is that, in this case, if a packet becomes active during a control slot, then the probability of a control failure is zero (i.e., it only reduces the probability of a control failure).

This then yields \lemref{noinject}, which shows that if no reset occurs during a streak, then the probability of a control failure during the streak is bounded.  As this is simply a union bound over the control slots, again, nothing changes.

\paragraph{Bounding resets.}
We then look at resets.  \lemref{trialsuccesses} is a standard statement of probability, unrelated to the protocol at hand.  

We then show \lemref{epochrestart}, which argues that a reset occurs during an epoch with at most constant probability, specifically, $1- (q/2)$ for some constant $q>0$ depending only on $\dd$.

Here, we need to modify the proof slightly.  Consider a slot-group with an empty control slot.  In this case, there are two cases that lead to us ``counting'' an empty data slot (and increasing the likilihood of reset):
\begin{itemize}
\item The data slot is empty.
\item The data slot is non-empty, followed immediately by an empty data slot.  
\end{itemize}
The first non-empty data slot triggers another data slot (for synchronization purposes), which may gives a second chance for an empty data slot.  (Recall that we count a data slot as empty for a slot-group in a three-slot slot-group if the second data slot is empty.)

Thus, we modify the proof as follows.  First, we consider epochs whose initial contention is 32.  This ensures that through the epoch, the contention is at least 2, and hence the probability of an empty data slot is at most $e^{-2d}$ (by \lemref{nonempty}).  

As such, the probability that, for a given slot-group, we count a data slot as empty is at most $2e^{-2d}$.  
%
Hence, for the purpose of \lemref{epochrestart}, we define $p = 1-2e^{-d}$.  The remainder of the analysis continues as before, with the conclusion that at least $p/4 \geq 1-\gamma$ fraction of the slots in the epoch are full with probability at least $q/2$.

\paragraph{Successful streaks.}
We proceed to consider the probability that streaks and epochs are successful.  \lemref{sigmastreak} follows from the definition of a streak, and is unchanged.

\lemref{consecutivestreaks} is a critical lemma that shows that every streak is successful with constant probability.  This follows from a calculation based on the probability of control failures and resets.  As those bounds remain unchanged, the lemma holds unchanged here.
As a result, \corref{epochsgood} also holds unchanged, where lengths refer to the number of slot-groups (instead of the number of slots).

\paragraph{Bad borrower games.}
The next part of the proof defines a combinatorial game, the bad borrower game.  This is independent of the protocol, and hence \lemref{borrowerbound}, \lemref{badborrower}, and \lemref{infiniteBBgame} hold unchanged.

\corref{infiniteEpochThroughput} then shows that in an infinite execution, if all the interstitial slots are removed, then we get good throughput.  This follows directly from the infinite bad borrwer game, and hence remains true.

\paragraph{Interstitial slots.}
We proceed to analyze the interstitial slots.  \lemref{packetprefix} shows that for any prefix of a packet's lifetime (as long as it is length at least 2), at least a $(1-\gamma)$ fraction of the data slots are full.  This remains true, so long as we count only the second data slots in a 3-slot slot-group.

\lemref{boundempty} then follows, implying that for timeslots that are non-quiet arrivals, at most a $\gamma$-fraction of the data slots are empty.  Again, this holds identically, so long as we only count the second data slot in a 3-slot slot-group.  (Moreover, the first data-slot in a 3-slot slot-group is always non-empty, by definition.)  

At this point, we have accounted for all the empty interstitial slots.  We now need to show that we get good throughput in the non-empty interstitial slots, which follows from the fact that a good broadcast is more likely than a collision.  This follows from calculating, in \lemref{collisionSuccess} the probability of a broadcast and a collision in a data slot, which remains unchanged.  It would not hold in a slot in which a new packet became active---but then a new epoch would begin (and it would no longer be an interstial slot).

From this we conclude \lemref{fullFiniteInterstitial} excactly as before, i.e., there are at most $O(n)$ full non-disrupted interstitial slots.  Putting the pieces together, this yields \thmref{finite} which shows expected constant throughput in finite executions.

The analysis of interstitial slots in the infinite case proceeds similarly, with \lemref{infinitethroughput} following as before from the same basic analysis of random walks (which does not depend on the protocol itself, and hence is unchanged).

\subsection*{Energy}

It remains to consider the energy usage of the modified protocol.  There are essentially no changes to the analysis, beyond accounting for the energy used in the extra data slots.

As before, a packet has a constant probability of terminating before it resets (\lemref{successBeforeReset}) because finishes as soon as it reaches a large enough stretch that is undisrupted.  

As in the general case, this yields \corref{fewresets}, i.e., that a packet has a probability at meast $1/e^{\Theta(k)}$ of resetting $k$ times.

We can then show \lemref{packetEnergy}, i.e., that a packet uses at most $O(\log^2(t))$ energy in its first $t$ slots.  Notice that a packet may now spend somewhat more energy, as it may now broadcast in two data slots for some slot-groups instead of just one.  This change, however, increases the energy usage by at most a constant factor.

From this, we get, in the finite case, that the total energy is bounded by $\log^2(n+\Cost)$, where $\Cost$ is the number of disrupted slots (\lemref{finiteEnergy}).  Similarly, in the infinite case, we get \lemref{infiniteEnergy} which bounds the energy in an infinite execution.

\end{document}